\documentclass[twocolumn]{emulateapj}
\usepackage{longtable}
\usepackage{graphics}
\usepackage{rotating}
\usepackage{amsmath}
\usepackage{txfonts}
\usepackage{natbib}
\usepackage{hyperref}

\def\gtorder{\mathrel{\raise.3ex\hbox{$>$}\mkern-14mu
             \lower0.6ex\hbox{$\sim$}}}
\def\ltorder{\mathrel{\raise.3ex\hbox{$<$}\mkern-14mu
             \lower0.6ex\hbox{$\sim$}}}

\slugcomment{\today}

\shorttitle{Supernova iPTF\,17cw}

\shortauthors{Corsi et al.}

\begin{document}

\title{\lowercase{i}PTF17cw: An engine-driven supernova candidate discovered independent of a gamma-ray trigger}
\author{A.~Corsi\altaffilmark{1},
  S.~B.~Cenko\altaffilmark{2,3},
  M.~M.~Kasliwal\altaffilmark{4},
  R.~Quimby\altaffilmark{5,6},
  S.~R. Kulkarni\altaffilmark{4},
  D.~A.~Frail\altaffilmark{7},
  A.~M.~Goldstein\altaffilmark{8},
  N. ~Blagorodnova\altaffilmark{4},
  V.~Connaughton\altaffilmark{8},
  D.~A. Perley\altaffilmark{9},
 L.~P.~Singer\altaffilmark{2},
  C.~M. Copperwheat\altaffilmark{9},
   C.~Fremling\altaffilmark{10},
 T.~Kupfer\altaffilmark{4},
 A.~S. Piascik\altaffilmark{9},
I.~A. Steele\altaffilmark{9},
F.~Taddia\altaffilmark{10},
H.~Vedantham\altaffilmark{4},
A.~Kutyrev\altaffilmark{2,11},
N.~T.~Palliyaguru\altaffilmark{1},
O.~Roberts\altaffilmark{12,13},
J.~Sollerman\altaffilmark{10},
E.~Troja\altaffilmark{2,11},
S.~Veilleux\altaffilmark{3,11}.}
 \altaffiltext{1}{Department of Physics and Astronomy, Texas Tech University, Box 1051, Lubbock, TX 79409-1051, USA; e-mail: alessandra.corsi@ttu.edu}
 \altaffiltext{2}{Astrophysics Science Division, NASA Goddard Space Flight Center, Greenbelt, MD 20771, USA.}
 \altaffiltext{3}{Joint Space-Science Institute, University of Maryland, College Park, MD 20742, USA,} 
\altaffiltext{4}{Division of Physics, Mathematics, and Astronomy, California Institute of Technology, Pasadena, CA 91125, USA.}
\altaffiltext{5}{Department of Astronomy / Mount Laguna Observatory, San Diego State University, San Diego, CA 92182, USA.}
\altaffiltext{6}{Kavli IPMU (WPI), UTIAS, The University of Tokyo, Kashiwa, Chiba 277-8583, Japan.}
\altaffiltext{7}{National Radio Astronomy Observatory, P.O. Box O, Socorro, NM 87801, USA.}
\altaffiltext{8}{Universities Space Research Association, NSSTC, 320 Sparkman Drive, Huntsville, AL 35805, USA.}
\altaffiltext{9}{Astrophysics Research Institute, Liverpool John Moores University, IC2, Liverpool Science Park, 146 Brownlow Hill, Liverpool, L3 5RF, UK.}
\altaffiltext{10}{Oskar Klein Centre, Department of Astronomy, Stockholm University, Albanova University Centre, SE-106 91 Stockholm, Sweden.}
\altaffiltext{11}{Department of Astronomy, University of Maryland, College Park, MD 20742, USA.}
\altaffiltext{12}{NASA Postdoctoral Fellow.}
\altaffiltext{13}{Astrophysics Office, ZP12, NASA/Marshall Space Flight Center, Huntsville, AL 35812, USA.}
\begin{abstract}
We present the discovery, classification, and radio-to-X-ray follow-up observations
of iPTF17cw, a broad-lined (BL) type Ic supernova (SN) discovered by the intermediate Palomar Transient Factory (iPTF). Although 
unrelated to the gravitational wave trigger, this SN was discovered as a happy by-product of the extensive observational 
campaign dedicated to the follow-up of Advanced LIGO event GW170104.
The spectroscopic properties and inferred peak bolometric luminosity of iPTF17cw are most similar to the gamma-ray burst (GRB) associated SN\,1998bw, while the shape of the $r$-band 
light curve is most similar to that of the relativistic SN\,2009bb. 
Karl G. Jansky Very Large Array (VLA) observations of the iPTF17cw field reveal a radio counterpart $\approx 10$ times less luminous than SN\,1998bw, and with peak radio luminosity comparable to that of SN\,2006aj/GRB\,060218 and SN\,2010bh/GRB\,100316D. 
 Our radio observations of iPTF17cw imply a relativistically expanding outflow. However, further late-time observations with the VLA in its most extended configuration 
are needed to confirm fading of iPTF radio counterpart at all frequencies. X-ray observations carried out with \textit{Chandra} reveal the 
presence of an X-ray counterpart with luminosity similar to that of SN\,2010bh/GRB\,100316D. Searching the \textit{Fermi} catalog for possible 
$\gamma$-rays reveals that  GRB\,161228B is spatially and temporally compatible with iPTF17cw. The similarity to SN\,1998bw and SN\,2009bb, the radio and 
X-ray detections, and the potential association with GRB\,161228B, all point to iPTF17cw 
being a  new candidate member of the rare sample of optically-discovered engine-driven BL-Ic SNe associated with relativistic ejecta. 
\end{abstract}

\section{Introduction}
\label{Introduction}
Almost 20 years after the discovery of an association between the radio-loud, broad-lined (BL) supernova (SN) of type Ic SN\,1998bw, and the low-luminosity, long GRB\,980425 \citep{Galama1998,Kulkarni1998}, we still have only a handful of 
GRB-associated (or engine-driven) SNe with radio light curves \citep{Berger030329,Soderberg2004,Frail2005,Soderberg060218,Margutti2013,Perley2014,Singer2015}. While the link between long GRBs and BL-Ic SNe is well established \citep{Bloom2006}, it remains unclear what makes some BL-Ic SNe launch ultra-relativistic  jets (GRBs).  Radio observations are key to shed light on this open question since radio probes the fastest moving ejecta and is unaffected by beaming  
effects that preclude detection of X-rays and $\gamma$-rays from off-axis GRBs that may drive some BL-Ic SNe. 

The discovery of explosions such as SN\,2009bb \citep{Soderberg2010}, a relativistic 
BL-Ic SN without a detected GRB, has also opened the intriguing question of whether there exists a population of events with properties in between that of ``ordinary'' BL-Ic SNe and GRBs (ultra-relativistic jets).  Radio, 
by being effective at probing also mildly relativistic explosions, can be used as a tool to search for these events ``in the gap'' \citep[see also][]{Ofek2007,Milisavljevic2015}.

Over the last decade, the above considerations have motivated extensive searches for radio emission from type Ib/c SNe \citep[e.g.,][]{Berger2003,Soderberg2006,Bietenholz2013}, which have constrained the fraction of Ib/c SNe associated with GRBs to $\lesssim 1-3\%$. Most of these studies, however, have long been limited by the very small number of BL-Ic SNe available to the community \citep{Berger2003,ATEL2483,Soderberg2010,ATEL3101,Drake2013,ATEL4997,Salas2013,Cha2015,Milisavljevic2015}. BL-Ic SNe are the only type of SNe observationally linked to GRBs \citep{Bloom2006}, but not all BL-Ic SNe are engine-driven and in fact the number of BL-Ic SNe with relativistic ejecta discovered independently of a $\gamma$-ray trigger remains extremely small.

Recently, we have begun to make progress toward carrying out a systematic study of BL-Ic SNe in the radio \citep{Corsi2011,Corsi2014,Corsi2016}, thanks to the much-increased rate of BL-Ic discoveries enabled by the Palomar Transient Factory \citep[PTF and its successor, the iPTF;][]{Law2009,Rau2009}. Over the course of 5 years, we have greatly enlarged the sample of BL-Ic SNe (discovered independently of a GRB trigger) with radio follow-up, and constrained the fraction of BL-Ic SNe as relativistic as (and observationally similar to) SN\,1998bw to $\lesssim 41\%$ of the BL-Ic population \citep[99.865\% confidence;][]{Corsi2016}. This upper-limit is starting to approach the lower-limit of $\gtrsim 30\%$ for the fraction of engine-driven BL-Ic SNe set by earlier studies by comparing BL-Ic SNe and GRB rates \citep{Podsiadlowski2004}. 

\begin{figure}
\begin{center}
\includegraphics[width=8.5cm]{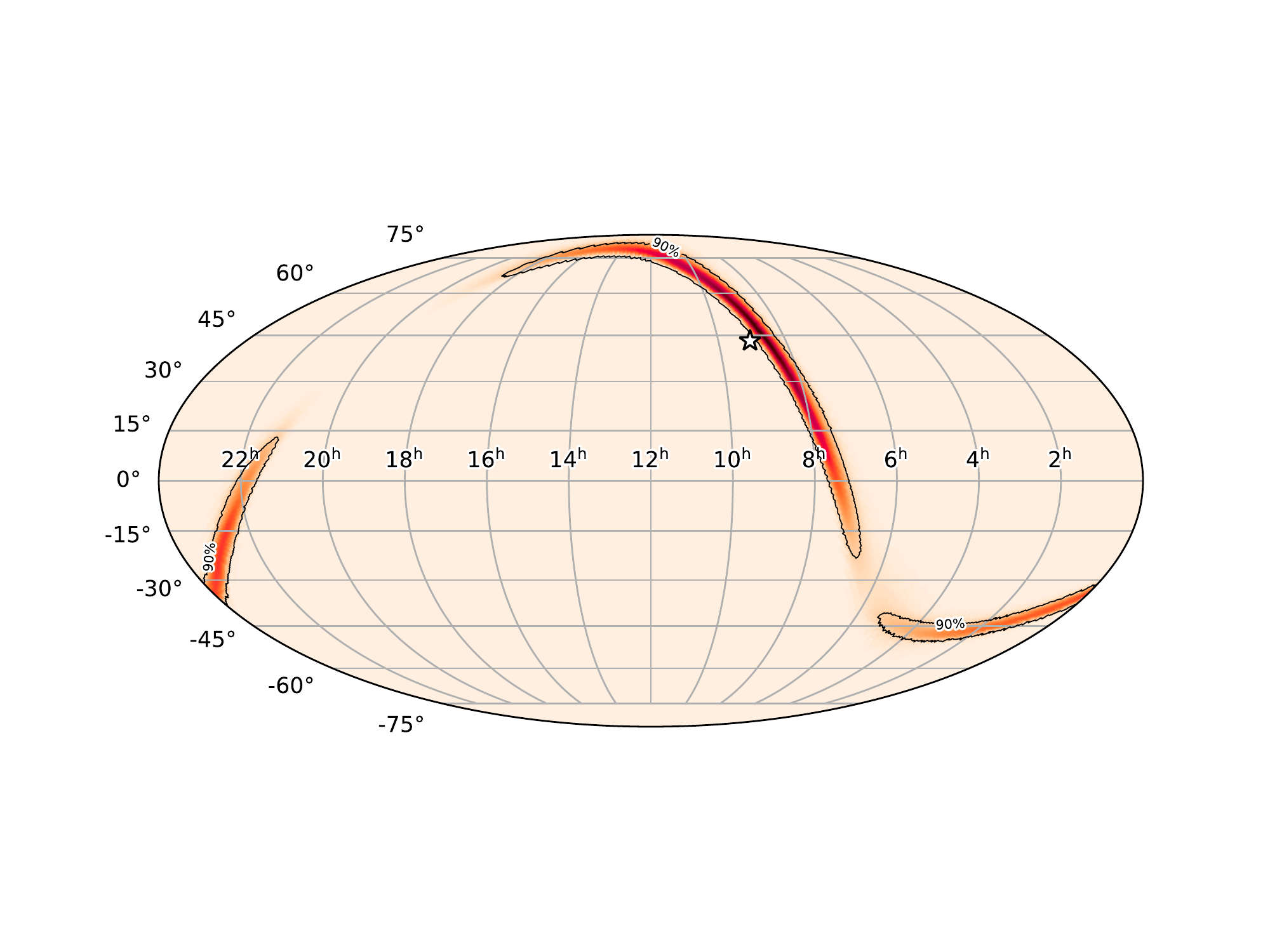}
\vspace{-1.0cm}
\caption{Position of iPTF17cw superimposed on the final LIGO localization \citep{LVC21056}. Color represents probability density. A black contour is drawn around the LIGO 90\% credible region. Because of the binary-black hole nature of GW170104 \citep{Abbott2017}, and given that iPTF17cw falls just outside of the $90\%$ localization area of GW170104, we rule out any association between iPTF17cw and GW170104.  \label{bayes}}
\end{center}
\end{figure}

Assuming that relativistic BL-Ic SNe are of order $30-40\%$ of the total, one would expect to have a high probability of discovering at least one relativistic BL-Ic SN in a sample of $\approx 17-22$ BL-Ic SNe with good radio follow-up. The available sample of BL-Ic SNe with radio observations has now reached this size \citep{Corsi2016}, and here we present our first discovery for a candidate relativistic BL-Ic SN, dubbed  iPTF17cw. This event was discovered by the iPTF during an extensive follow-up campaign of the Advanced LIGO event GW170104 \citep{LVC20364,LVC20385,LVC20398,LVC20419,LVC20428,LVC20473,LVC21056}. Because of the binary-black hole nature of GW170104 \citep{Abbott2017}, and given that iPTF17cw falls just outside of the $90\%$ localization area of GW170104 (Fig. \ref{bayes}), we rule out any association between iPTF17cw and GW170104.  

In what follows, we describe our dataset (Section \ref{Observations}), analyze our multi-wavelength observations (Section \ref{Analysis}), and show that iPTF17cw adds one more candidate to the sample of  BL-Ic SNe from which we have ever detected radio emission (indicative of relativistic expansion) independent of a $\gamma$-ray trigger. iPTF17cw also adds one more event to the more general class of relativistic optical transients discovered independently of a high-energy trigger \citep{Cenko2013,Cenko2015}. An after-the-fact search for $\gamma$-rays reveals that iPTF17cw may be associated with GRB\,161228B. In our conclusion (Section \ref{conclusion}), we stress how further late-time radio observations of the iPTF17cw field are needed to securely confirm the relativistic nature of iPTF17cw.  

\section{Panchromatic observations}
\label{Observations}
\subsection{SN optical photometry}
\label{Opticalphotometry}
\begin{figure}
\begin{center}
\includegraphics[width=8.6cm]{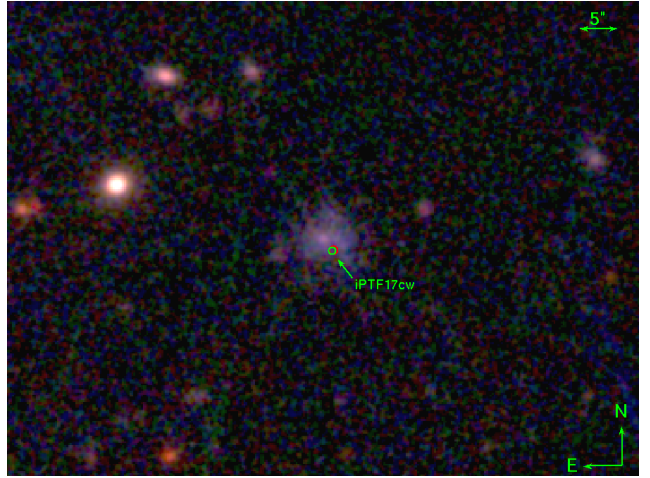}
\vspace{-0.1cm}
\caption{Pan-STARRS 1 \citep{Kaiser2010,PS1} pre-imaging of the host galaxy of iPTF\,17cw, with the
location of the optical transient superimposed (0.5$\arcsec$ error circle shown in green, which also contains the position of the radio counterpart).  We also show the error circle of the X-ray counterpart to iPTF17cw (red). The host is blue, extended, and diffuse, with a
reddish central concentration.  The optical position (as well as radio and X-ray positions) are significantly offset (by $\approx 1.7\arcsec$) from the center of the host galaxy. \label{fig:host}}
\end{center}
\end{figure}

We discovered iPTF17cw on 2017 January 7 UT in an $R$-band image taken with the 7.8\,deg$^2$ wide camera on the Palomar Samuel Oschin 48-inch Schmidt Telescope (P48), which is routinely used by the intermediate Palomar Transient Factory \citep{Cao2016,Law2009,Masci2017}.  iPTF17cw is located at $\alpha=09^{\rm h}03^{\rm m}38^{\rm s}.38$ and $\delta=+43^{\circ}05'50''.3$ (J2000; Fig. \ref{fig:host}), at an angular separation of $\approx 1.8''$ from the nominal position of SDSS\,J090338.47+430551.6 \citep{York2000}. 

At the time of discovery, iPTF17cw was visible at a magnitude of $R\approx 19.5$\,mag. The SN was not visible in a previous $g$-band image of the field obtained on 2016 December 6 with the P48. Subsequent observations were performed with the Palomar 60-inch Telescope \citep[P60; ][]{Cenko2006} in $gri$ bands. Photometry was  performed  relative  to  the  SDSS magnitudes of stars in the field, using our custom pipelines \citep{Ofek2012,Fremling2016}. 

\begin{figure}
\hspace{-0.5cm}
\includegraphics[width=9cm]{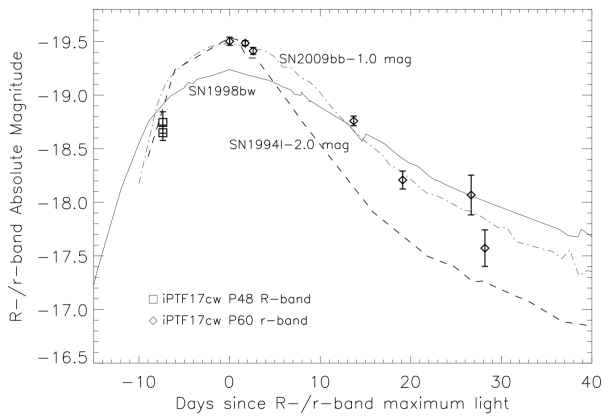}
\vspace{-0.35cm}
\caption{The $P48$ $R$-band and $P60$ $r$-band light curves of iPTF17cw (corrected for galactic extinction) are compared to the (extinction-corrected) $R$-band light curves of: the GRB-associated BL-Ic SN\,1998bw \citep[solid;][]{Clocchiatti2011}; the relativistic BL-Ic SN\,2009bb \citep[dash-dotted;][]{Pignata2011}; the type Ic SN\,1994I \citep[dashed;][]{Richmond1996}. Both the SN\,2009bb and SN\,1994I light curves have been scaled to match the peak luminosity of iPTF17cw. \\\label{fig:phot}}
\end{figure}

In Figure \ref{fig:phot} we show the P48 $R$-band and P60 $r$-band light curves of iPTF17cw. In Table \ref{phot_tab} we report the multi-color photometry of iPTF17cw.

\begin{figure*}
\begin{center}
\hbox{
\hspace{-0.4cm}
\includegraphics[width=9.5cm]{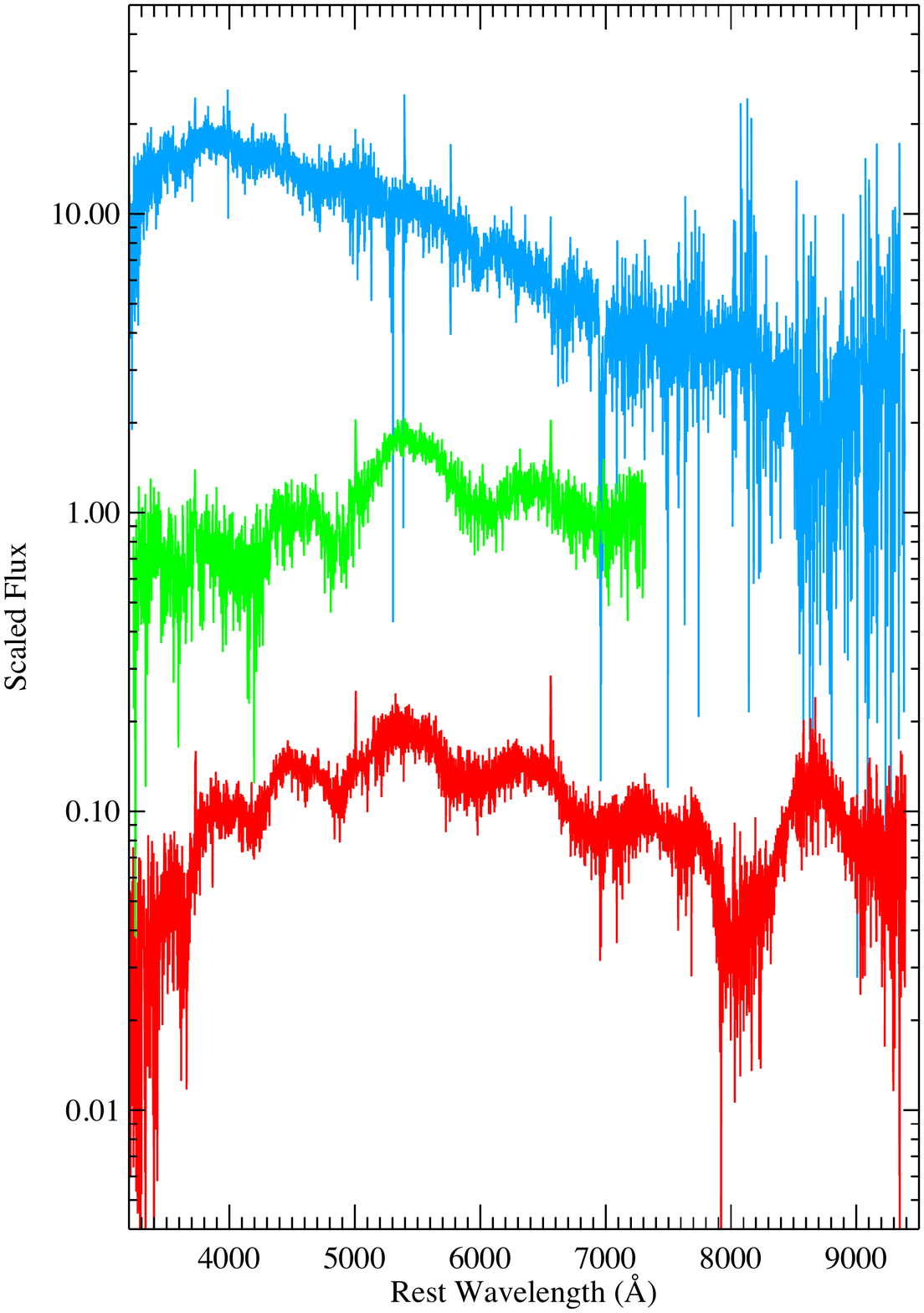}
\hspace{-1cm}
\includegraphics[width=9.5cm]{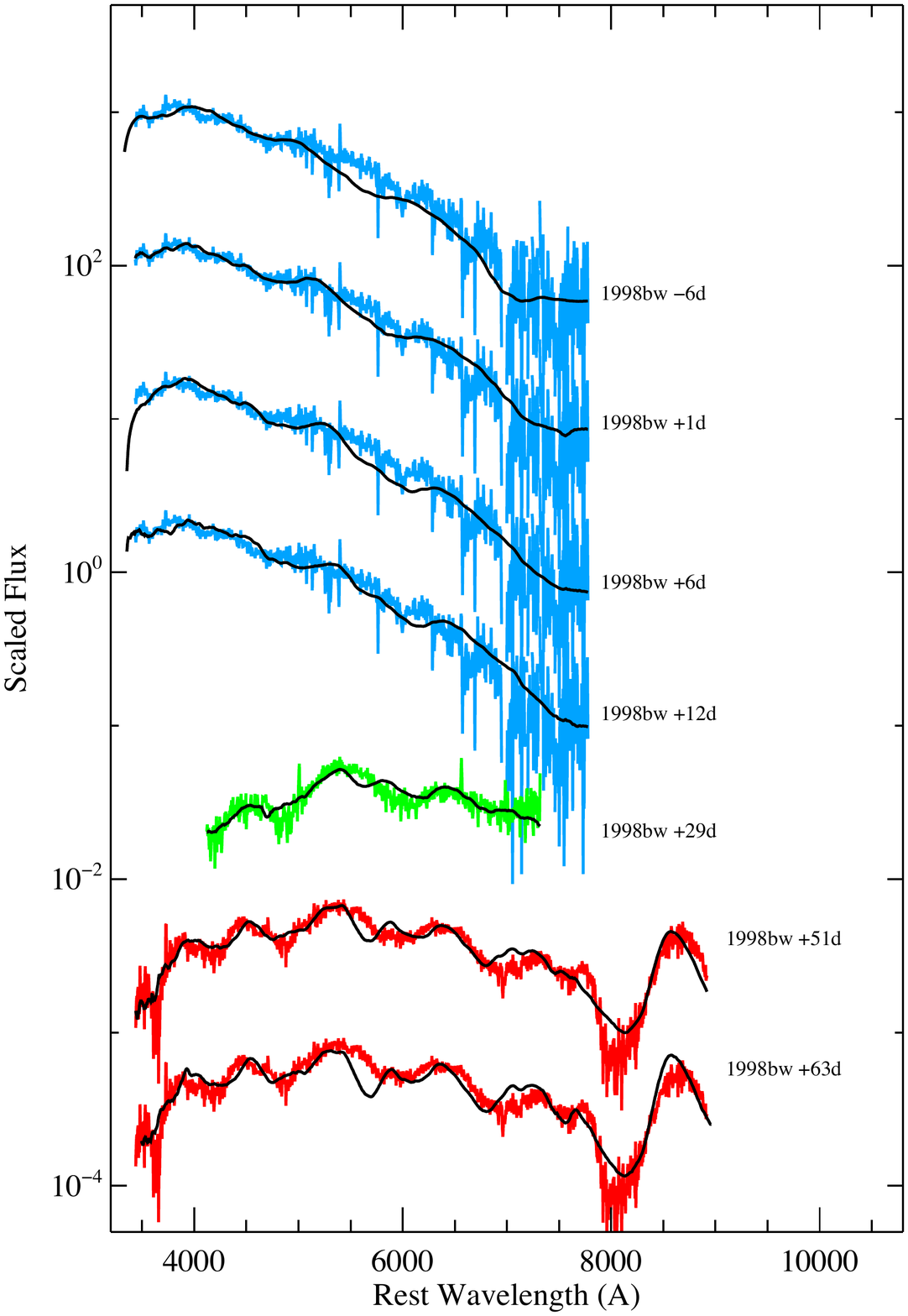}}
\caption{Left:  iPTF17cw spectra (blue for P200/DBSP, $\approx -8$\,d since $r$- and $g$-band maximum light; green for DCT, $\approx +17$\,d since $r$- and $g$-band maximum light; red for Keck/LRIS, $\approx +43$\,d since $r$- and $g$-band maximum light), corrected for redshift effects assuming $z=0.093$. Right: same iPTF spectra as in the left panel, but binned and galaxy template subtracted, compared to de-reddened SN\,1998bw spectra at various phases (see text for discussion). \label{fig:spec}}
\end{center}
\end{figure*}

\begin{figure}
\begin{center}
\hspace{-0.5cm}
\includegraphics[width=7cm,angle=-90]{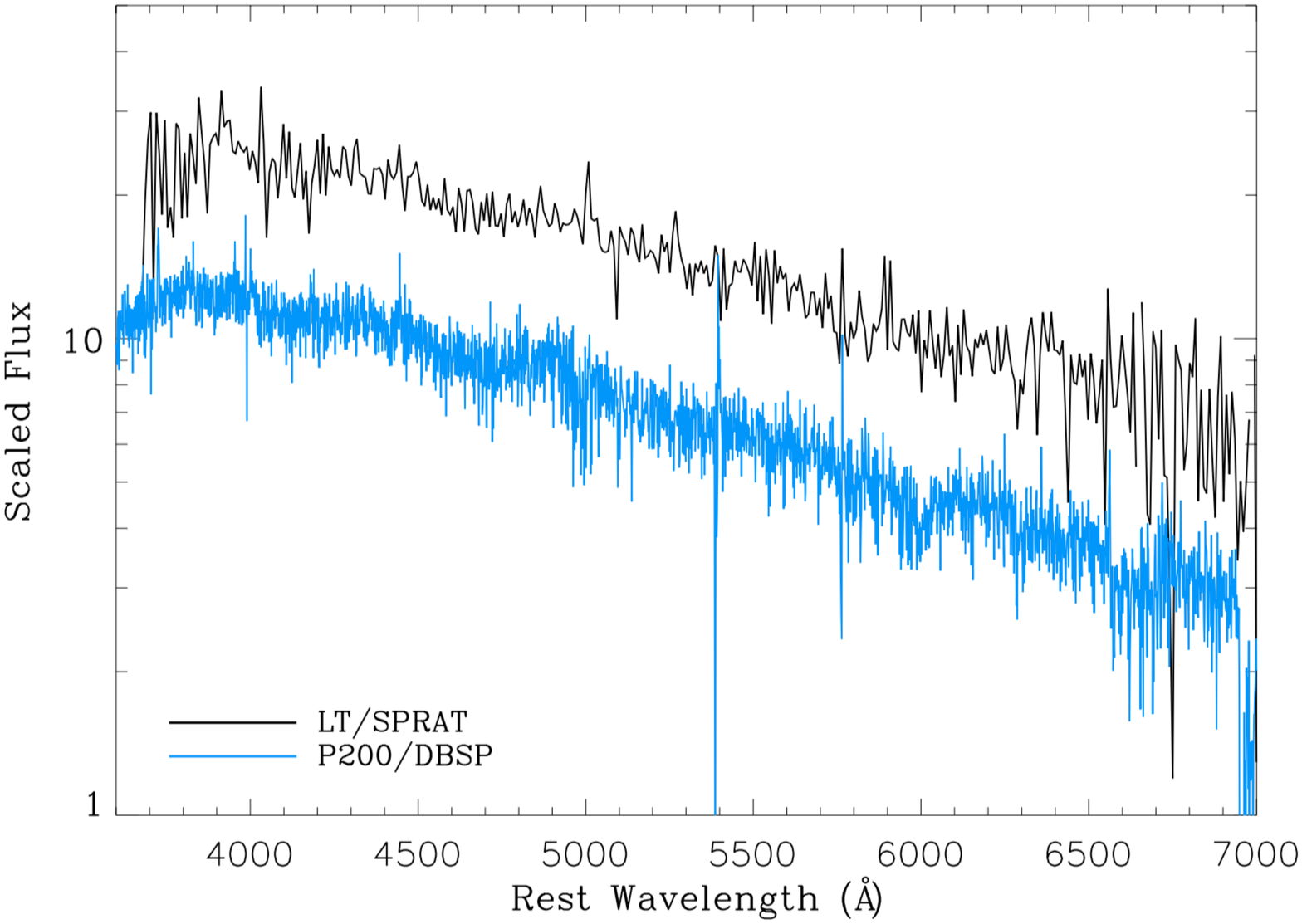}
\caption{LT/SPRAT spectrum of iPTF17cw (at $\approx -6$\,d since $r$-band maximum light) compared to the P200/DBSP spectrum (at $\approx -8$\,d since $r$-band maximum light).  \label{fig:LT}}
\end{center}
\end{figure}

\subsection{Spectral classification}
\label{Spectralclassification}
A summary of our spectroscopic observations of iPTF17cw is reported in Table \ref{tab:optspec}. 

We observed iPTF17cw with the Double Spectrograph \citep[DBSP;][]{Oke1982} on the
Palomar 5.1\,m Hale telescope for 600\,s beginning at 13:20\,UT on
2017 January 7 (MJD 57760.555; Fig. \ref{fig:spec}, blue). The blue channel used the 600 lines/mm grating
giving an effective wavelength coverage of 3085\,\AA\ to 5840\,\AA. 
The red channel used the 316 lines/mm grating, which
covers the range 5200\,\AA\ to 10900\,\AA. The D55 dichroic was used
to divide light into the two channels (50\% transmission at 5500\,\AA). 
The observations used a 1\,\arcsec slit mask oriented close to the
parallactic angle \citep{f82}. The DBSP data were processed and optimally extracted using a custom
pipeline implemented in IRAF\footnote{IRAF is distributed by the National Optical Astronomy
Observatory, which is operated by the Association of Universities for
Research in Astronomy (AURA) under a cooperative agreement with the
National Science Foundation.}, python,
and IDL. Background light was fit for and removed from the 2-D frames
using the IDL routine {\tt bspline\_itterfit.pro} following the
procedure described by \citet{kelson2003}. An initial wavelength
solution was found using calibration lamp observations. This solution
was then adjusted based on night sky lines to account for flexture in
the instrument. The flux scale was determined using observations of
the standard star Feige 34 taken on the same night.

Narrow emission lines are detected at 4073\,\AA, 5471\,\AA, and
7171\,\AA. Identifying these as O~II $\lambda\lambda$\,3726,3728, O~III
5007, and H$\alpha$ from the host galaxy, we determine a redshift of
$z=0.093$. Aside from these narrow features, the spectra show
subtle, broad features superposed on a continuum that peaks around
3800\,\AA\ in the rest frame (Fig. \ref{fig:spec}, left, blue line). Using the spectral template fitting
code, {\tt superfit} \citep{howell2006}, we find that the spectrum is
reasonably well matched to the BL-Ic SN\,1998bw around maximum light \citep[Fig. \ref{fig:spec}, right; see also][]{Iwamoto1998}.

An additional spectrum of iPTF17cw  was obtained by combining three spectra collected around a similar epoch (2017 January 9 UT) using the 
SPRAT spectrograph \citep{SPRAT} on the 2\,m robotic Liverpool Telescope. SPRAT is a 
long slit spectrograph with a wavelength range from 4000-8000\,\AA\, and a resolving power of $R\sim 350$ 
at the centre of this range. These data were reduced, wavelength and flux calibrated using the automated 
Liverpool Telescope spectroscopic pipeline, and the combined spectrum was obtained using STARLINK/FIGARO software \citep{Currie2014}. The SPRAT spectrum of iPTF17cw is consistent with the above described P200 spectrum collected at a similar epoch (see Figure \ref{fig:LT}).

\begin{center}
\begin{longtable}{lllll}
\caption{Spectroscopic observations of iPTF17cw. \\Phases are in days since \textit{observed} $g$- and $r$-band maximum light. \\All spectra will be made available via  WISeREP \citep{Yaron2012}.  \label{tab:optspec}}
\\
\hline
\hline
Date & Phase & Instrument & Wavelength range & Exp. Time \\
(MJD) & (d) &  & (\AA)& (s) \\ 
\hline
\endhead
57760.555 & $\approx -8$ &P200/DBSP &  3085-10900 & 600 \\
55762.249 & $\approx -6$ & LT/SPRAT & 4000-8000 & $3\times600$\\
57785.358 & $\approx +17$ &DCT/DeVeny &  3500-8000 & 600 \\
57811.454 &  $\approx +43$ &  Keck I/LRIS & 3150-9400 & 1800 / 1700\\
\hline
\end{longtable}
\end{center}

We continued monitoring of iPTF17cw with the DeVeny spectrograph mounted on the 4.3\,m
Discovery Channel Telescope (DCT) in Happy Jack, Arizona.  We obtained
a series of 4 $\times$ 600\,s exposures beginning at 8:36 UT on 1
February 2017 (MJD 57785.358).  We employed the 300 lines mm$^{-1}$ grating with a 
1.5\,\AA\ slit, resulting in a spectral resolution of $\approx 10$\,\AA\ and
wavelength coverage from $\approx 3500$--8000\,\AA.  All spectra 
were obtained at the parallactic angle \citep{f82} and
reduced in the IRAF environment using standard routines.  Spectra were
extracted optimally \citep{h86}, and we performed wavelength calibration
first relative to HgNeArCd arc lamps and then tweaked based on night
sky lines.  Extracted spectra were divided through by a smoothed
flux standard to remove narrowband ($< 50$\,\AA) instrumental effects
\citep{b99}.  Finally, telluric atmospheric absorption features were
removed using the continuum from spectrophotometric standards
\citep{wh88}.  The resulting spectrum is shown in Figure \ref{fig:spec} (green).

A last spectrum of iPTF17cw was taken with Keck I/LRIS instrument \citep{Oke1995} at 10:53:39 UT on 2017 February  27  (MJD 57811.454). 
The instrument was configured in long slit mode, with $1\arcsec$ slit and 400/3400+400/8500 grism+grating combination 
(dispersion of 1.09 and 1.16\,\AA/pixel, respectively). The total exposure time was 1800\,s in the blue channel and 1700\,s in the red. 
The resolution of the spectrum, as measured from the O I 5577\,\AA\ line is of 9.6\,\AA\ ($\approx 440$\,km/s).
The data were reduced using a custom developed pipeline in IDL, LPIPE. The standard stars Feige34 and Feige67 
were used for flux calibration and correction for telluric lines. This spectrum is show in Figure \ref{fig:spec} (red).

\subsection{Radio follow-up}
\label{Radioobservations}
Radio observations of the iPTF17cw field were carried out with the Karl G. Jansky Very Large Array\footnote{The National Radio Astronomy Observatory is a facility of the National Science Foundation 
operated under cooperative agreement by Associated Universities, Inc.} (VLA) 
in its A, AnD, and D configurations, under our target of opportunity program\footnote{VLA/16B-043; PI: A. Corsi.}. 
Our first observation was carried out in C-band (nominal central frequency of $\approx 6$\,GHz) on 2017 January 13 
(approximately 6\,d since $P48$ discovery). VLA follow-up observations were continued for up to about three months 
since optical discovery. We used J0920+4441 as our phase calibrator, 3C48 or 3C286 as flux / bandpass calibrators.
VLA data were reduced and calibrated using using the VLA automated calibration pipeline which runs in the Common 
Astronomy Software Applications package \citep[CASA;][]{McMullin2007}. When necessary, additional flags were applied after visual inspection of the data. 
Images of the observed field were formed using the CLEAN algorithm \citep{Hogbom1974}, which we ran in 
interactive mode. The results of our VLA follow-up campaign of iPTF17cw are reported in Table \ref{radioTab}. Flux measurement errors are calculated as the quadratic sum of the map rms, plus a 5\% fractional error on the measured flux which accounts for inaccuracies of
the flux density calibration.

\begin{figure}
\vbox{
\includegraphics[width=9cm]{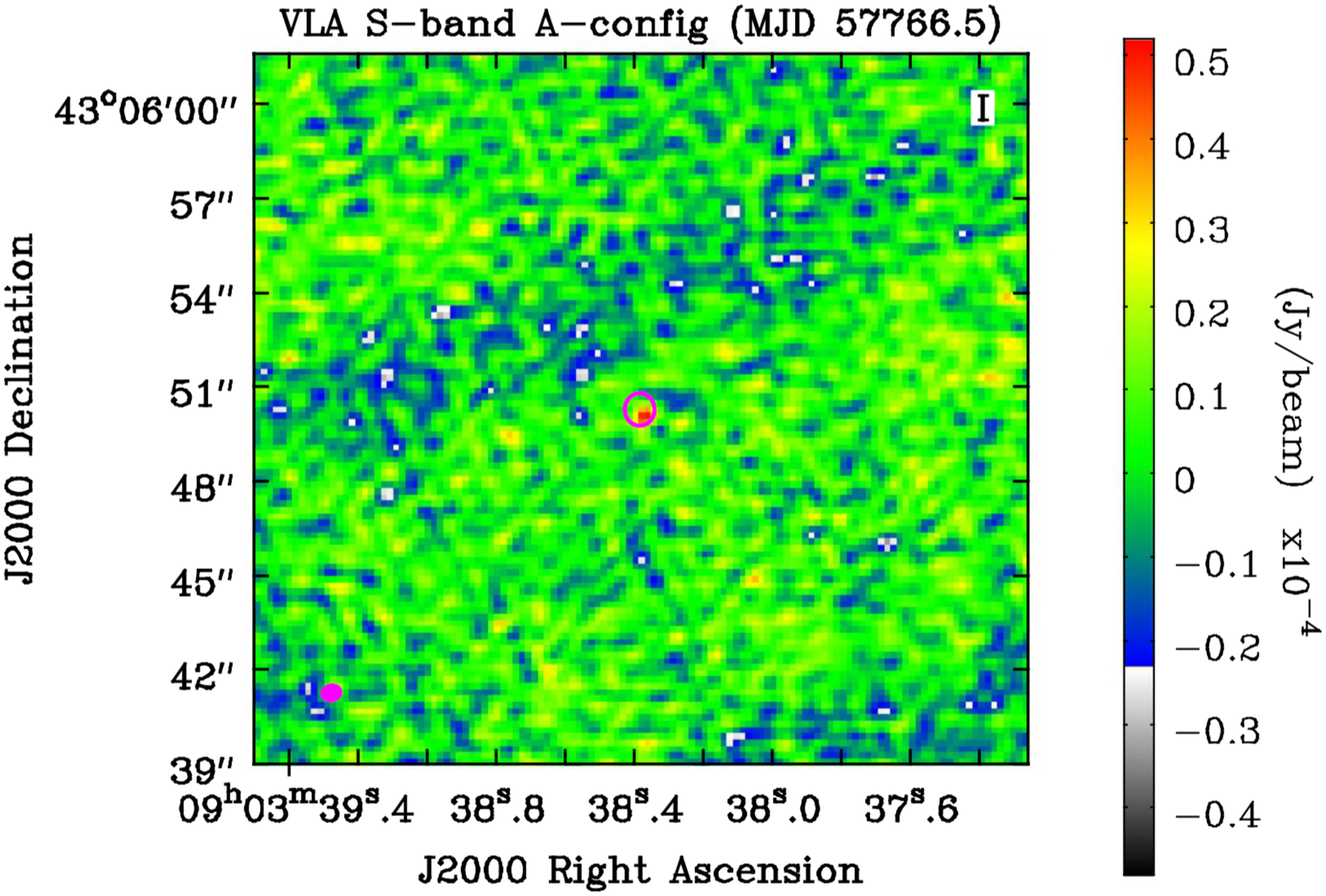}
\includegraphics[width=9.cm]{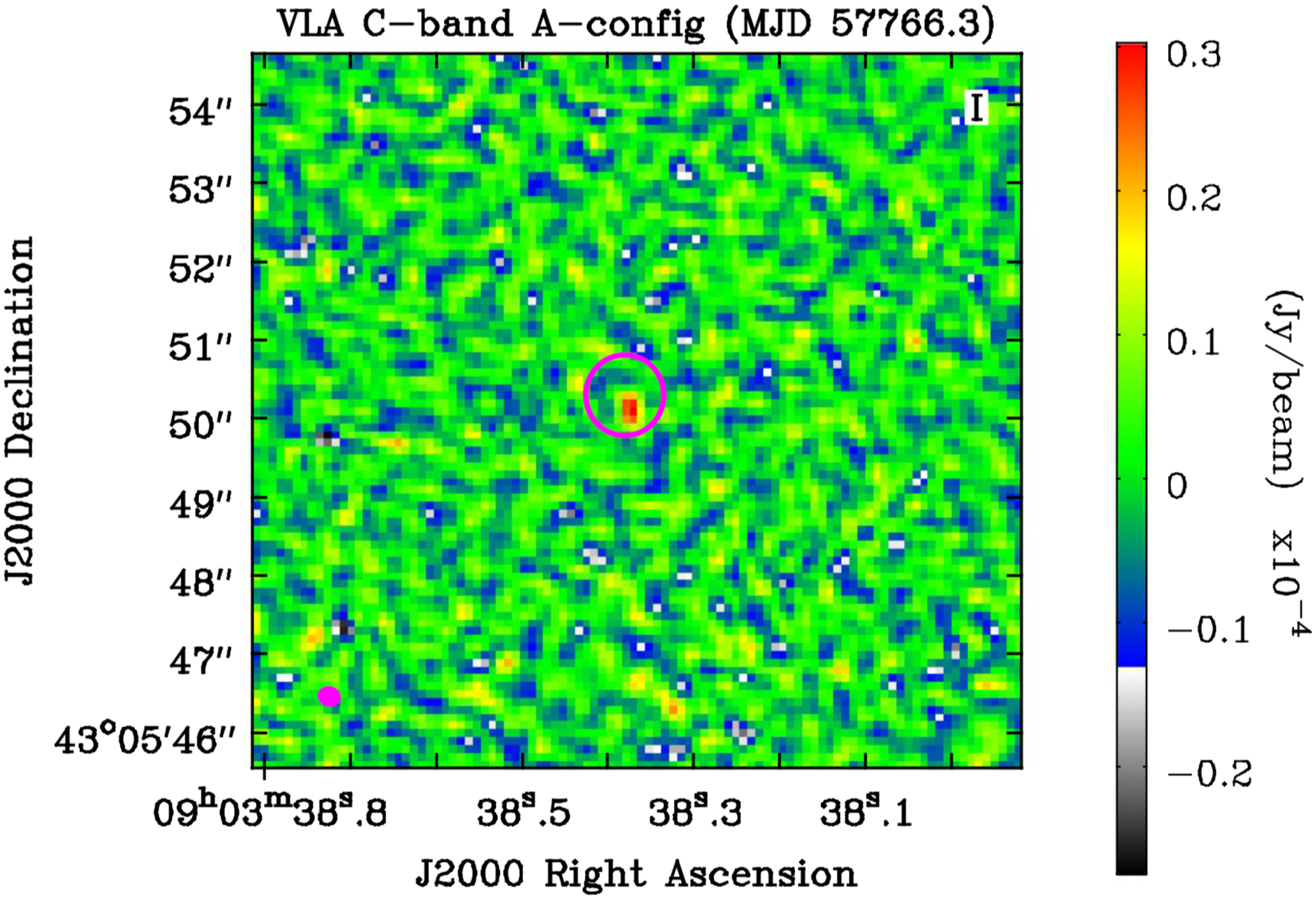}
\includegraphics[width=9.3cm]{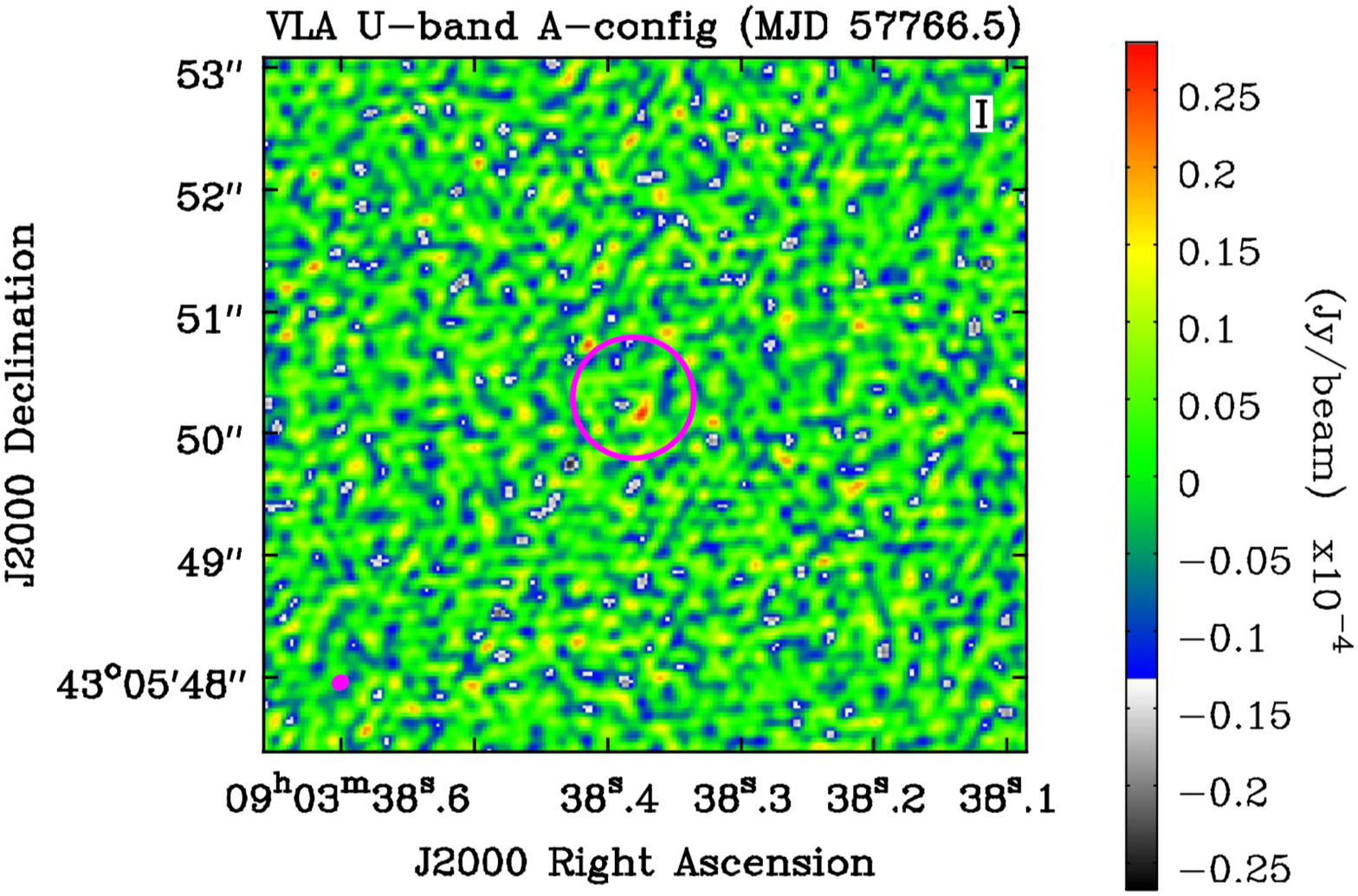}}
\caption{Radio images (2.8\,GHz, top; 6\,GHz, middle; 14\,GHz, bottom) of the iPTF17cw field as seen by the VLA in its A configuration. The optical position of iPTF17cw is at the center of the 0.5\arcsec-radius magenta circle (iPTF localization). The size of the VLA beam at each frequency is shown as a magenta filled ellipse in the bottom-left corner of each panel. The radio counterpart of iPTF17cw is within the iPTF localization circle and appears to be point-like at all frequencies (as evident by comparing its size with the size of the synthesized VLA beam in magenta). \label{fig:radiodiscovery}}
\end{figure}

\begin{figure}
\vbox{
\includegraphics[width=8cm]{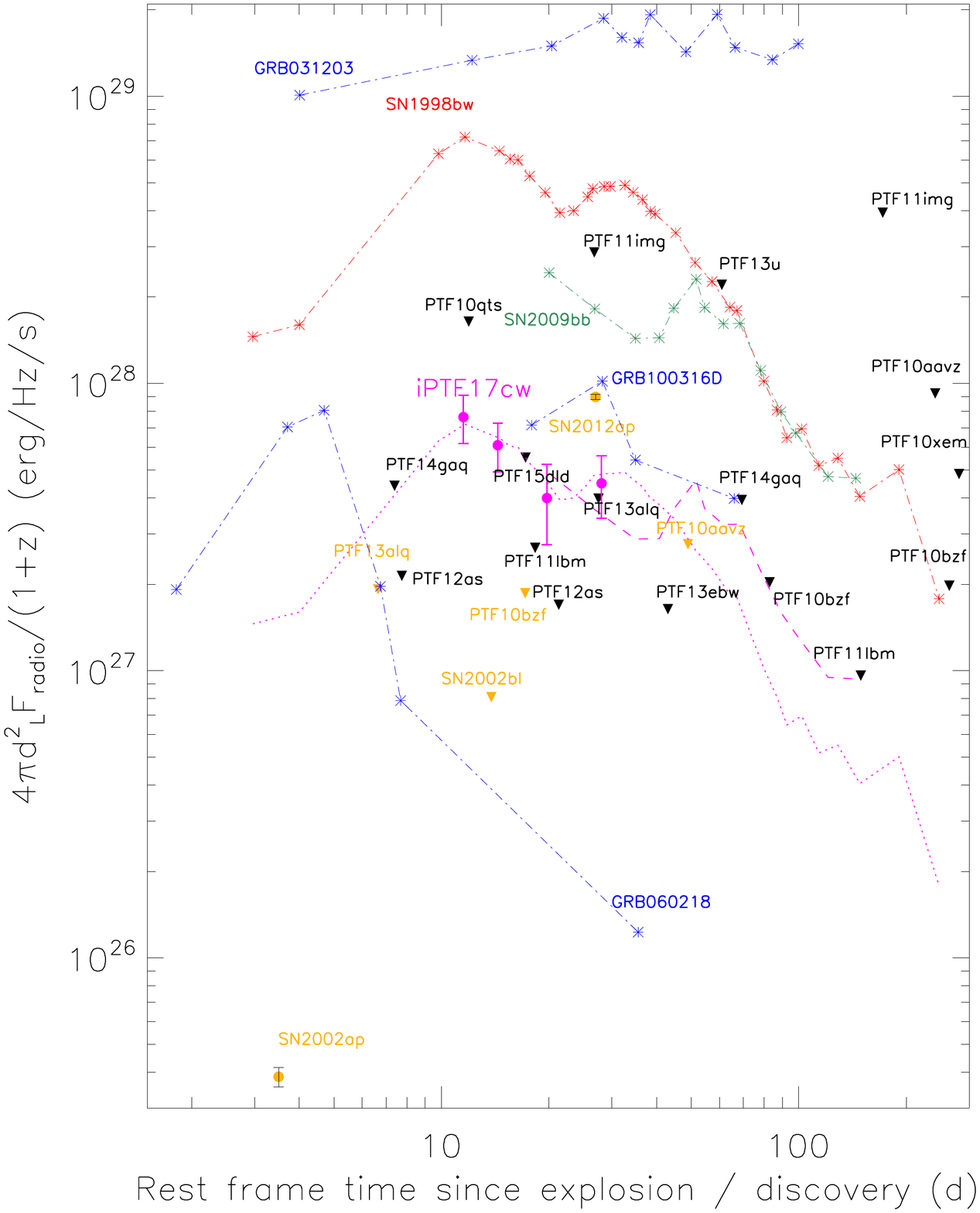}
\includegraphics[width=8.6cm]{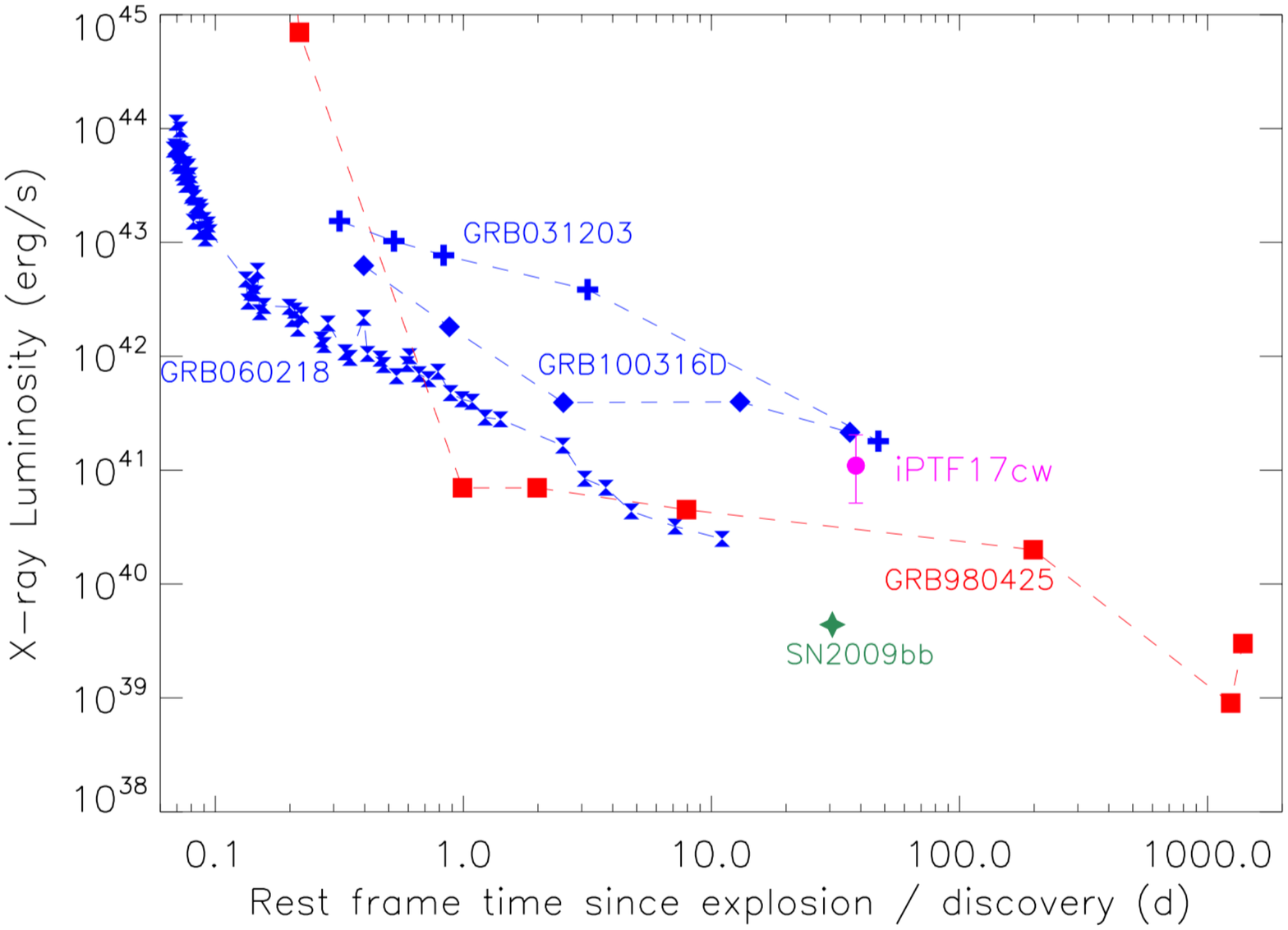}
}
\caption{Top: Radio ($\approx 5$\,GHz) light curves of some low-luminosity GRBs \citep[blue;][]{Soderberg2004,Soderberg2006,Margutti2013}, of the GRB-associated BL-Ic SN\,1998bw \citep[red;][]{Kulkarni1998}, the relativistic BL-Ic SN\,2009bb \citep[green;][]{Soderberg2010} and SN\,2012ap \citep[yellow dot;][]{Cha2015}, and the non-relativistic BL-Ic SN\,2002ap \citep[yellow dot;][]{Berger2003}, are compared to the $6$\,GHz radio emission from iPTF17cw (magenta dots). The light curves of SN\,1998bw and SN\,2009bb scaled so as to match the radio luminosity of iPTF17cw are also shown for comparison (magenta dotted and dashed lines, respectively). We also show upper-limits for the sample of BL-Ic SNe without a radio detection collected as part of our (black downward pointing triangles) or other (yellow downward pointing triangles) programs. Here we do not show BL-Ic SNe in our sample with radio detections due to strong CSM interaction \citep{Corsi2014,Corsi2016}. See text for discussion. Bottom: X-ray luminosity of some of the low-luminosity GRBs \citep{Kouveliotou2004,Watson2004,Campana2006,Margutti2013} and relativistic SNe \citep{Soderberg2010} plotted in the top panel, compared to iPTF17cw. See text for discussion. \label{fig:radio}}
\end{figure}

 A faint radio source was 
detected at a location compatible with the optical position of iPTF17cw, at a $\approx 5\sigma$ level 
during our first observation at $\approx 6$\,GHz. The source appears to be point-like with the VLA in its A configuration (nominal FWHP at 6 GHz of $\approx 0.33$\arcsec), thus a contribution from star formation in the underlying host galaxy is unlikely. The positional offset with respect to the reddish central concentration of the iPTF17cw host galaxy (Fig. \ref{fig:host}) also makes a contribution from a low-level AGN unlikely. In Figure \ref{fig:radiodiscovery} we show the VLA images of the iPTF17cw radio counterpart collected during our first multi-band observation of this source.

Over the first three C-band epochs, which cover a time interval of $\approx 10$ days, the source radio flux at $\approx 6$\,GHz 
shows marginal evidence ($\approx 2\sigma$ level) for variability suggesting a decay of the radio flux with time ($f_{\rm 6\,GHz}\propto (t-t_0)^{\alpha}$ with $\alpha =-1.18\pm0.68$ when $t_0$ is set equal to the trigger time of GRB\,161228B; see Section \ref{gammarays}), perhaps followed by what appears to 
be a flattening or re-brightening of the 6\,GHz emission (Fig. \ref{fig:radio}, top). Due to the change of the VLA configuration and to the presence of a 
bright QSO in the field,  our latest VLA image of the iPTF17cw field at 6\,GHz is substantially degraded (the image is dynamic range limited) and resulted in a non detection. 

The candidate radio 
counterpart to iPTF17cw was also detected at the $\gtrsim 3\sigma$ level at 2.8\,GHz and 14\,GHz. The observations at such frequencies do not show
evidence for substantial variability over the first two months of observations, but our last observation at 14\,GHz shows that the radio counterpart to iPTF17cw has faded below our detection level.

\subsection{X-ray observations}
\label{X-rayobservations}
The \textit{Swift} satellite \citep{gcg+04} observed the location of 
iPTF17cw on two epochs, beginning at 13:48 UT on 13 January
2017, and 00:35 UT on 18 January 2017.  These observations were 
obtained as part of our approved target of opportunity program\footnote{\textit{Swift} Cycle 12 GI; PI: A. Corsi}. 
We analyzed the data obtained
by the on-board X-Ray Telescope \citep[XRT;][]{bhn+05} during these
two epochs (1.4 and 7.5\,ks of exposure time, respectively) using
the procedures outlined in \citet{ebp+09}.  No X-ray emission was
detected at the location of iPTF17cw.  We place $(3\sigma)$ upper limits on the
0.3--10\,keV count rate of $< 8.1 \times 10^{-3}$ and $< 2.0 \times
10^{-3}$\,count\,s$^{-1}$ in these two epochs.  For a power-law spectrum
with a photon index of $\Gamma = 2$ and a Galactic HI column density of 
$N_H\approx 1.8\times10^{20}$\,cm$^{-2}$ \citep{Dickey1990}, these correspond to (3$\sigma$)
limits on the 0.3--10\,keV (unabsorbed) X-ray flux of $< 3.4 \times 10^{-13}$
and $< 8.5 \times 10^{-14}$\,erg\,cm$^{-2}$\,s$^{-1}$, respectively.

We also observed the location of iPTF17cw with the Advanced CCD Imaging
Spectrometer (ACIS) on the \textit{Chandra} X-ray Observatory\footnote{Program ID: \#17508570; PI: 
Corsi} beginning at 9:22 UT on 8 February 2017.  The source location was positioned on the S3 chip and the 
detector data was telemetered in ``faint'' mode.  An effective
exposure time of 9.86\,ks was obtained on source. We detect a total of 5 photons within 1\arcsec\ (radius) of the 
position of iPTF17cw.  While not sufficient for any spectral 
analysis, we can claim a detection at the $\approx 3\sigma$ confidence level (expected number of background counts: 0.89).

The position of the iPTF17cw X-ray counterpart as seen by \textit{Chandra} is $\alpha=09^{\rm h}03^{\rm m}38^{\rm s}.37$ and $\delta=+43^{\circ}05'50''.4$ (J2000), with an estimated (90\% confidence) 
error radius of $\approx 0.6\arcsec$ (dominated by \textit{Chandra} pointing errors). This position is compatible with both the optical position of iPTF17cw, and the position of its radio counterpart (see Fig. \ref{fig:host}).

Properly accounting for the on-axis point-spread function we
measure a 0.5--7.0\,keV count rate of $5.0_{-2.9}^{+4.8} \times
10^{-4}$\,count\,s$^{-1}$ (90\% confidence interval).  Adopting a
power-law spectral model with photon index $\Gamma = 2$ and 
$N_H\approx 1.8\times10^{20}$\,cm$^{-2}$ \citep{Dickey1990}, this
corresponds to an unabsorbed X-ray flux in this bandpass of
$f_{X} = \left(5.0_{-2.9}^{+4.7}\right) \times 10^{-15}$\,erg\,cm$^{-2}$\,s$^{-1}$, or an X-ray luminosity of $\approx (4-19)\times10^{40}$\,erg\,s$^{-1}$  (Fig. \ref{fig:radio}, bottom).

\subsection{Search for $\gamma$-rays}
\label{gammarays}
\begin{figure}
\begin{center}
\vspace{-1.5cm}
\includegraphics[width=9cm]{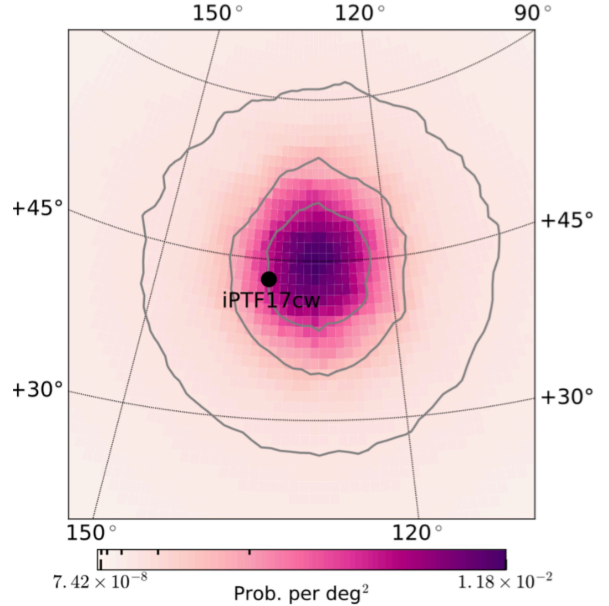}
\vspace{-2cm}
\caption{The GBM localization for GRB 161228B.  The three gray contours denote the 1-, 2-, and 3-$\sigma$ confidence regions, and the purple color gradient shows the probability per square degree for the true location of the source.  The location of iPTF17cw is shown to exist near the 1-$\sigma$ confidence contour.\label{localization}}
\end{center}
\end{figure}

We searched the \textit{Fermi} and \textit{Swift} catalogs for potential GRBs detected within a month prior to the iPTF discovery of iPTF17cw, and 
with position compatible with that of iPTF17cw. We found a match with GRB\,161228B (Fig. \ref{localization}), which was observed by \textit{Fermi}/GBM as trigger 
\#504623745\footnote{\url{https://gcn.gsfc.nasa.gov/other/504623745.fermi}} on 28 December 2016, at 13:15:41.85\,UT (MJD 57750.552). 
This GRB was also detected  by POLAR and \textit{Integral} \citep{GCN20348}.

GRB\,161228B, as observed by \textit{Fermi}/GBM, appears to be approximately 30\,s in duration.  Utilizing the position of 
iPTF17cw for the GBM detector responses, we performed a time-integrated spectral analysis starting at about 1\,s before the 
GBM trigger time and extending to 30\,s after trigger time.  The data are well fit by a Band function \citep[defined as in][]{Gruber2014} with the following parameter 
values: $E_{\rm peak}= 195\pm47$\,keV; low-energy power-law index, $\alpha=-1.2\pm0.1$; and high-energy power-law index, 
$\beta=-2.3\pm0.5$.  The derived fluence from this spectral fit in the typical GBM reporting band of 10-1000 keV is $(5.6\pm0.4)\times10^{-6}$\,erg\,cm$^{-2}$.  
Additionally, we analyzed the brightest 1 s interval of data for this GRB to estimate the peak flux.  The high-energy power-law index 
of the Band function could not be adequately constrained for this interval, however, a power-law with an exponential cutoff could be 
reasonably constrained with parameter values of $E_{\rm peak} = 214 \pm 26$\,keV and power-law index, $\alpha=0.3\pm0.3$.  The 1\,s peak flux is 
then estimated to be $(5.4\pm0.9)\times10^{-7}$\,erg\,s$^{-1}$\,cm$^{-2}$.

\section{Multi-wavelength analysis}
\label{Analysis}
\subsection{Photometric properties}
\label{opticalproperties}
To compare the intrinsic peak luminosity of iPTF17cw with that of other GRB-associated SNe, we use our $gri$ photometry from P60 and the prescriptions of \citet{Lyman2014} for transforming optical light curves of stripped-envelope core-collapse SNe to full bolometric light curves. We define the full bolometric correction in $g$-band as \citep{Lyman2014}:
\begin{equation}
BC_{g,\rm peak}=M_{\rm bol,peak}-M_{g,\rm peak},
\label{mbol_eq}
\end{equation}
where 
\begin{equation}
M_{\rm bol,peak}=M_{\odot,\rm bol}-2.5\log_{10}\left(\frac{L_{\rm bol,peak}}{L_{\odot,\rm bol}}\right).
\end{equation}
In the above relation, $M_{\odot,\rm bol}=4.74$\,mag and $L_{\odot,\rm bol}=3.828\times10^{33}$\,erg\,s$^{-1}$. 

Following \citet{Lyman2014}, we express $BC_g$ as a function of $g-i$ or $g-r$ colors. For stripped-envelope core-collapse SNe, the best-fit relations between colors and bolometric corrections give:
\begin{eqnarray}
 BC_{g}=0.054-0.195\times(g-r)-0.719\times(g-r)^2 \label{bolcor_eq1},\\
BC_{g}=-0.029-0.404\times(g-i)-0.230\times(g-i)^2 \label{bolcor_eq2}.
\end{eqnarray}

To derive iPTF17cw color, we correct our photometry for Galactic extinction toward the position of iPTF17cw \citep[$A_V=0.0517$\,mag which, assuming an extinction to reddening ratio $A_{V} / E(B-V) = 3.1$, implies $A_g=0.064$\,mag, $A_r=0.044$\,mag, $A_i=0.033$\,mag; ][]{Schlafly2011}. This yields $gri$-band extinction-corrected magnitudes of iPTF17cw at the time of $g$-band peak of $g\approx 18.98$\,mag, $r\approx 18.66$\,mag, and $i\approx 18.87$\,mag, respectively (see Table \ref{phot_tab}).  Using these into Equations (\ref{bolcor_eq1})-(\ref{bolcor_eq2}), we get two values of the bolometric correction at peak that averaged out yield $BC_{g,\rm peak}\approx -0.08$\,mag. We stress that this is a tentative estimate of the bolometric correction based on the results of \citet{Lyman2014} from best-fit relations between colors and bolometric corrections for a population of stripped-envelope core-collapse SNe. 

From Equation (\ref{mbol_eq}) we derive $M_{\rm bol,peak}\approx-19.26\,{\rm mag}$ which implies $L_{\rm bol,peak}\approx 10^{43}$\,erg\,s$^{-1}$. This is comparable to the bolometric luminosity of the engine-driven SN\,1998bw, for which $L_{\rm bol,peak}\approx (7.7\pm8.6)\times10^{42}$\,erg\,s$^{-1}$ \citep[see Table 3 in][]{Clocchiatti2011}. Applying k-corrections and host extinction corrections based on colors, we get that iPTF17cw is, in term of optical brightness, among the top $\approx 10\%$ of the iPTF BL-Ic SN sample \citep{Taddia2017}.

From the bolometric peak luminosity we can estimate the amount of $^{56}$Ni in the explosion, via the relation \citep[e.g.,][]{Arnett1982,Lyman2016}:
\begin{equation}
\log_{10}(M_{\rm Ni}/M_{\odot})=-0.415\times M_{\rm bol,peak}-8.184,
\end{equation}
which yields $M_{\rm Ni}\approx 0.64\,M_{\odot}$ for iPTF17cw, comparable to the literature values range of $(0.4-0.7)\,M_{\odot}$ for SN\,1998bw, and somewhat higher than $M_{\rm Ni}\approx 0.25\,M_{\odot}$ estimated for SN\,2009bb \citep[see Table 5 in][]{Lyman2016}. 

As evident from Figure \ref{fig:phot}, the $r$-band light curve shape of iPTF17cw is more similar to that of SN\,2009bb \citep{Pignata2011} than to that of SN\,1998bw. The light curve shape is related to the characteristic  timescale scale \citep[e.g.,][]{Arnett1982,Lyman2016}: 
\begin{equation}
\tau_m\approx 8\,{\rm d}\left(\frac{M_{\rm eje}}{M_{\odot}}\right)^{3/4}\left(\frac{E_K}{10^{51}\,\rm erg}\right)^{-1/4},
\end{equation}
where $M_{\rm eje}$ is the ejecta mass, $E_K$ is the kinetic energy, and where we have assumed on opacity of $\approx 0.06-0.07$\,cm$^2$\,g$^{-1}$. For SN\,2009bb,  $\tau_{m,\rm bb}\approx 10$\,d \citep{Lyman2016}, and given the similarity with the light curve of iPTF17cw, we derive for the last:
\begin{equation}
\left(\frac{E_K}{10^{51}\,\rm erg}\right)\approx0.4\times\left(\frac{M_{\rm eje}}{M_{\odot}}\right)^{3}.
\label{paraeq1}
\end{equation}

\subsection{Spectroscopic properties}

In Figure \ref{fig:spec} (right) we show the DBSP spectrum of iPTF17cw (see Section \ref{Spectralclassification}) collected on 
MJD 57760.555 (at a phase of $\approx -8$\,d since $r$- and $g$-band maximum light), compared to SN\,1998bw spectra 
at several phases since maximum light. The SN\,1998bw spectra shown in this Figure have been ``de-reddened''
significantly to match the iPTF17cw spectrum and to account for the much bluer continuum of iPTF17cw at early times.
The iPTF17cw DBSP spectrum was corrected for host galaxy contribution 
(a host galaxy light template was estimated using Superfit and then subtracted to the iPTF17cw spectrum). A comparison of the spectra 
shown in the left and right panels of Figure \ref{fig:spec} gives an idea of the effects of these corrections.

In Figure \ref{fig:spec} (right) we also show the DCT and Keck/LRIS spectra of iPTF17cw taken on MJD\,57785.358 ($\approx +17$\,d after $r$- and $g$-band maximum light; see Section \ref{Spectralclassification}) and on MJD\,57811.454 ($\approx +43$\,d after $r$- and $g$-band maximum light; see Section \ref{Spectralclassification}), respectively, compared to SN\,1998bw. Also in this case we have subtracted host galaxy light from the iPTF17cw spectrum. The amount of de-reddening used to match the SN\,1998bw spectra to that of iPTF17cw is significantly less than what needed for the earlier comparison described above, thus indicating that iPTF17cw had colors more similar to SN\,1998bw at this later phase. 

Based on the above comparisons, we can reasonably estimate the iPTF17cw photospheric velocity around maximum light to be similar to that of SN\,1998bw, $(19,500^{+1700}_{-1000})$\,km\,s$^{-1}$ \citep[see ][and references therein]{Lyman2016}. We note that if we assume that the minimum observed around 6600\,\AA~ in the DCT spectrum is due to the Si 6355\,\AA\ line, the estimated velocity  is of $(15,600\pm2000)$\,km\,s$^{-1}$ at $\approx +17$ since maximum light.  

We thus set \citep[see e.g. Eq. 2 in][]{Lyman2016}:
\begin{equation}
{\rm v}_{\rm phot,peak}\approx 19.5\times10^{3}{\rm \,km\,s}^{-1}=\left(\frac{10E_K}{3 M_{\rm eje}}\right)^{1/2},
\end{equation}
which implies:
\begin{equation}
\left(\frac{E_K}{10^{51}\,\rm erg}\right)\approx 2 \times\left(\frac{M_{\rm eje}}{M_{\odot}}\right).
\label{paraeq2}
\end{equation}
Combining the above Equation with equation (\ref{paraeq1}), we get for iPTF17cw:
\begin{eqnarray}
M_{\rm eje}\approx 2 M_{\odot},\\
E_K\approx 4 \times 10^{51}\,{\rm erg}.
\label{eneK}
\end{eqnarray}
These values are comparable to $M_{\rm eje}\approx 1.9\,M_{\odot}$ and $E_K\approx 3.3\times10^{51}$\,erg found for SN\,2009bb \citep[see Table 5 in][]{Lyman2016}, while SN\,1998bw had a larger ejecta mass and kinetic energy ($M_{\rm eje}\approx 4.4\,M_{\odot}$ and $E_K\approx 9.9\times10^{51}$\,erg).

\subsection{Association with GRB\,161228B}
The results from the spectral fits of GRB\,161228B described in Section \ref{gammarays} imply that GRB\.161228B is a ``typical'' GRB in terms of  Band function parameters of GBM-detected long 
GRBs, with values of $E_{\rm peak}$, peak flux, and fluence in the $\approx 50-60$\,th percentile \citep{Gruber2014}.  If we assume that GRB\,161228B and iPTF17cw are indeed associated, as suggested by the compatibility of localizations and trigger/discovery time (see Section \ref{gammarays}), we can use the redshift of iPTF17cw to calculate the k-corrected 
energy release and peak luminosity of GRB\,161228B over the 1\,keV-10 MeV band.  The derived isotropic-equivalent energy release in $\gamma$-rays, $E_{\rm iso}$, is $(2.3\pm0.6)\times10^{50}$\,erg and the peak luminosity, $L_{\rm peak}$, is $(1.8\pm0.2)\times10^{49}$\,erg\,s$^{-1}$.  These values 
are at least one order of magnitude lower than the average observed $E_{\rm iso}$ and $L_{\rm peak}$ values of long GRBs.  In fact, we 
estimate that these values are within the 1st percentile of long GRBs observed by GBM \citep{Gruber2014}.  Although we find that the energy and luminosity are not 
as low as for GRB\,980425, our results for GRB\,161228B are consistent with the observations of other type SN-associated GRBs such as GRB\,031203 \citep{Soderberg2004} and GRB\,120422A  \citep{Melandri2012}. We stress, in particular, the similarity with GRB\,031203, which also appeared to be an event with spectrum similar to cosmological GRBs, but with a much lower isotropic energy release \citep{Sazonov2004}. 

We estimate the probability of a chance coincidence between GRB\,161228B and iPTF17cw by calculating the rate of \textit{Fermi} GRBs falling in a random $103\deg^2$ 
patch of sky (as large as the $1\sigma$ region of GRB\,161228B localization; see Fig. \ref{localization}). This  rate is $\approx 0.05$ per month, after taking into account the combined statistical and systematic localization uncertainties. Thus, a chance coincidence between iPTF17cw and GRB\,161228B within a month time window has a probability of $\approx 5\%$.

In order to further test the compatibility of GRB\,161228B explosion time with that of iPTF17cw, we fit the $R$-band light curve of SN\,1998bw to the (k-corrected) $R$-band light curve
of iPTF17cw \citep[for more details, see][]{Taddia2017}. In the fitting process, we allow the light curve  of SN\,1998bw to be stretched in time and shifted in magnitude, until we minimize the residuals of the difference between the stretched SN\,1998bw light curve and the light curve of iPTF17cw. This fitting procedure is similar to that outlined in \citet{Cano2013} and takes advantage of the known explosion epoch of SN\,1998bw. 
We obtain a good fit at an explosion epoch corresponding to $\approx$ MJD 57752, which is not incompatible with the trigger time of GRB\,161228B.

We conclude that the association between iPTF17cw and GRB161228B is plausible. In what follows, we thus set the explosion time of iPTF17cw to be the trigger time of GRB\,161228B (MJD 57750.552).

\subsection{Properties of the fastest (radio-emitting) ejecta}
\label{radiomodel}
\begin{figure}
\hspace{-0.3cm}
\includegraphics[width=8.3cm]{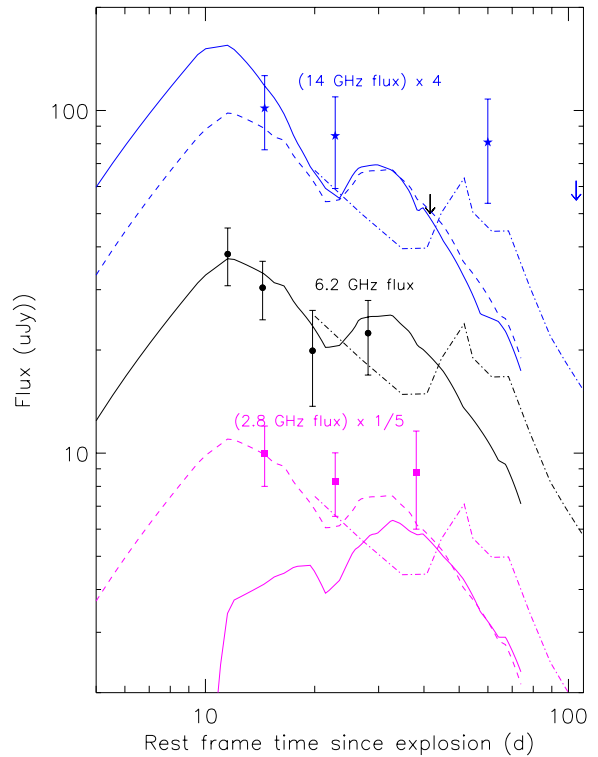}
\vspace{-0.2cm}
\caption{The multi-band (2.8-14 GHz) radio light curves of iPTF17cw are shown and compared with the (interpolated) light curves of SN\,1998bw at similar frequencies, scaled to match the luminosity of iPTF17cw (solid lines). The extrapolation of the scaled 5\,GHz light curve of SN\,1998bw to 2.8\,GHz (magenta dashed line) and 14 GHz (blue dashed line) using an optically thin spectral index of $\beta\approx -0.5$ are also shown. The 5\,GHz scaled light curve of SN\,2009bb, and its extrapolations to 2.8\,GHz and 14\,GHz using the same spectral index of $\beta\approx -0.5$, are shown with dash-dotted lines. See text for discussion.\label{fig:radiolight}}
\end{figure}

The temporal evolution of the candidate radio counterpart to iPTF17cw is shown in Figure \ref{fig:radio} and Figure \ref{fig:radiolight}. In these Figures, we set the time of explosion to the trigger time of GRB\,161228B. As evident from Figure \ref{fig:radio}, the 6\,GHz luminosity of iPTF17cw is $\approx 10\times$ smaller than GRB\,980425/SN\,1998bw, and $\approx 5\times$ smaller than the relativistic SN\,2009bb. The GHz radio luminosity of iPTF17cw is orders of magnitudes below the typical radio emission from cosmological GRBs \citep[e.g.,][]{Chandra2012}, but it is not incompatible with the radio emission observed from low-luminosity GRBs such as GRB\,060218/SN\,2006j and GRB\,100316D/SN\,2010bh. The radio spectral energy distribution suggests an optically thin spectrum around MJD 57766 ($\approx 16$\,d since GRB\,161228B). However, the measurements errors are large.  Using the VLA multi-band observations collected around MJD 57766, we estimate a 2.8-14\,GHz spectral index ($f_{\nu}\propto \nu^{\beta}$) of $\beta=-0.44\pm0.39$ around this epoch. 

In Figure \ref{fig:radiolight} we show the  2.8\,GHz to 14\,GHz radio light curves of iPTF17cw compared to the radio light curves of SN\,1998bw at similar frequencies (solid lines), all scaled by a factor of $\approx 10$ in intrinsic luminosity to match the lower intrinsic luminosity of iPTF17cw (see Fig. \ref{fig:radio}). Given the large distance to iPTF17cw, our radio measurements are affected by large errors. Thus, in what follows, we do not attempt a detailed modeling of iPTF17cw radio emission, but rather provide orders-of-magnitude estimates of the key properties of its radio emitting ejecta, based on comparisons with other radio SNe.

The 6\,GHz radio emission of iPTF17cw is qualitatively compatible with the double-peaked light curve of SN\,1998bw, within the large measurement errors. However, a slower-evolving or later-peaking second radio peak, perhaps more similar to the one observed for SN\,2009bb at 5\,GHz (dash-dotted lines), would be needed to account for the flat temporal behavior of the iPTF17cw counterpart at 14\,GHz. For what concerns the emission at 2.8\,GHz, we note that SN\,1998bw emission was suppressed by synchrotron self-absorption (SSA) at the lower frequencies until $t\approx 40$\,d since explosion (magenta solid line in Fig. \ref{fig:radiolight}). Indeed, if we extrapolate the (scaled) 5\,GHz light curve of SN\,1998bw to our other frequencies of interest (2.8\,GHz and 14\,GHz) using an optically thin spectral index of $\beta\approx -0.5$ (dashed lines in Fig. \ref{fig:radiolight}), we see that this extrapolation agrees well with the actual (scaled) light curves of SN\,1998bw at 14\,GHz (blue solid line), while at 2.8\,GHz the agreement is good only at $t\gtrsim 40$\,d (magenta dashed and solid lines). iPTF17cw radio emission, on the other hand, appears to be optically thin at all frequencies $\gtrsim 2.8$\,GHz already at $t\gtrsim 16$\,d since explosion. In fact, the 2.8\,GHz emission of iPTF17cw is compatible with the optically-thin extrapolation to 2.8\,GHz of the (scaled) 5 GHz light curve of SN\,1998bw (magenta dashed line in Fig. \ref{fig:radiolight}). This, together with the fact that iPTF17cw emission appears to decrease with time at 6.2\,GHz over the first three epochs, suggests that the SSA peak frequency is $\nu_p\lesssim 2.8$\,GHz at $\approx 16$\,d since explosion (GRB\,161228B), and thus $F_p \gtrsim 50\,\mu$Jy (or $L_p\approx 10^{28}$\,erg\,s$^{-1}$\,Hz$^{-1}$ at the distance of iPTF17cw).

The above constraints on the SSA peak frequency and flux enable us to estimate the size of the radio emitting material at $t\approx16$\,d since explosions. Using Eq. (11) of \citet{Chevalier1998}:
\begin{equation}
R_{p}\approx 6.6\times10^{15}\,{\rm cm}\left(\frac{\eta}{2\alpha}\right)^{1/17.2} \left(\frac{F_p}{\rm Jy}\right)^{8.1/17.2}\left(\frac{d_L}{\rm Mpc}\right)^{16.2/17.2}\left(\frac{\nu_p}{5\,\rm GHz}\right)^{-1},
\label{eq:ejspeed}
\end{equation}
where $\alpha \approx 1$ is the ratio of relativistic electron energy density to magnetic energy density, $F_{p}$  is the flux at the time of SSA peak, $\nu_{p}$ is the SSA frequency, and where we have parametrized with $\eta$ the thickness of the radiating electron shell and we assumed $p\approx 2.1$ for the power-law index of the electron energy distribution. Setting in the above Equation $F_p\gtrsim 50\,\mu$Jy and $\nu_{p}\lesssim 2.8$\,GHz, we get $ R_p\gtrsim 3.5\times10^{16}(\eta/5)^{1/17.2}$\,cm at 16\,d since explosion, which implies that the velocity of the radio ejecta is ${\rm v}_s\gtrsim[0.85\,c\,(\eta/5)^{1/17.2} ]$\footnote{A more precise calculation for iPTF17cw radio ejecta speed is obtained by considering the transverse shock speed constraint $\Gamma_s \beta_s c=R_p/t\approx 0.85\,c $, which implies $\Gamma_s\gtrsim 1.3$ or $\beta_s\gtrsim 0.65\,c$.}, similarly to what found for SN\,2009bb \citep{Soderberg2010}.

The magnetic field can be expressed as \citep[see Eq. (12) in][]{Chevalier1998}:
\begin{equation}
B_p\approx 0.5 \left(\frac{\eta}{2\alpha}\right)^{4/17.2}\left(\frac{F_{p}}{\rm Jy}\right)^{-2/17.2}\left(\frac{d_{L}}{\rm Mpc}\right)^{-4/17.2}\left(\frac{\nu_p}{5\,\rm GHz}\right){~\rm G},
\label{eq2}
\end{equation}
where we have set $p\approx 2.1$. The above gives $B_p\approx 0.3(\eta/5)^{4/17.2}$\,G at $t\approx 16$\,d, for $\alpha \approx 0.1$, $F_p\approx 50\,\mu$Jy, and $\nu_{p}\approx 2.8$\,GHz.

Hereafter, we continue our discussion of the properties of the radio-emitting region of iPTFcw using a simple radiation model that neglects the dynamics (evolution of the radius $R$, magnetic field $B$, and minimum Lorentz factor of the accelerated electrons with time), and relativistic effects related to the high ejecta velocity. Based on the analysis of SN\,1998bw by \citet{Li1999}, we expect  the corrections caused by dynamical and relativistic effects to be relatively modest, so that our estimates below are likely within a factor of $\lesssim 2$ of the values that would be derived via a more detailed modeling.

For a SN shock expanding in a circumstellar medium (CSM) of density:
\begin{equation}
\rho=5 \times10^{11} \,{\rm g\,cm}^{-1}A_{*}\times R^{-2}=\frac{\dot{M}/(10^{-5}M_{\odot}/{\rm yr})}{4\pi {\rm v}_w/(10^3{\rm km/s})}\times R^{-2},
\end{equation}
assuming that a fraction $\epsilon_B$ of the energy density $\rho{\rm v}^2_s$  goes into magnetic fields, one can write \citep[see Equation (8) in][]{Chevalier2006}:
\begin{equation}
 \frac{B_p^2}{8\pi}\approx 3\times10^{-2}A_* \left(\frac{\epsilon_B}{0.1}\right)\left(\frac{R_p}{3.5\times 10^{16}\,\rm cm}\right)^{-2}\left(\frac{{\rm v}_s}{0.85\,c}\right)^2\,{\rm erg\,cm^{-3}}.
\label{eq1}
\end{equation}
From the above we get $A_*\approx0.1  (\eta/5)^{8/17.2}$, or $\dot{M}\approx 10^{-6}(\eta/5)^{8/17.2}\,M_{\odot}{\rm yr}^{-1}$, for $\epsilon_B\approx 0.1$ and using the values of the shock radius and shock speed found previously. The total energy coupled to the fastest (radio emitting) ejecta around $t\approx 16$\,d can be expressed as:
\begin{equation}
E_r\approx \frac{4\pi R_p^3}{\eta}\frac{B_p^2}{8\pi\epsilon_B}=\frac{R_p^3}{\eta}\frac{B_p^2}{2\epsilon_B},
\label{eq3}
\end{equation}
which yields $E_r\approx 4\times10^{48}(\eta/5)^{11/17.2}$\,erg for $\epsilon_B\approx 0.1$ and using the value of the magnetic field found above. This energy is $\approx 0.1\%$ of the total kinetic energy of the ejecta estimated via optical photometry (see Equation \ref{eneK}). 

The values of energy and CSM density found for iPTF17cw can be compared with  $\dot{M}\approx2.5\times10^{-7}M_{\odot}{\rm yr}^{-1}$ and $E_r\approx (1-10)\times10^{49}$\,erg estimated for SN\,1998bw by \citep{Li1999}, with $\dot{M}\approx2\times10^{-6}M_{\odot}{\rm yr}^{-1}$ and $E_r\approx 1.3\times10^{49}$\,erg estimated for SN\,2009bb \citep{Soderberg2010}, and with  with $\dot{M}\approx(0.4-1)\times10^{-5}M_{\odot}{\rm yr}^{-1}$ and $E_r\approx (0.3-4)\times10^{49}$\,erg estimated for GRB\,100316D \citep{Margutti2013}. 

We finally note that because of the potential association with GRB\,161228B, the early-time evolution of the radio (and X-ray counterparts) to iPTF17cw may be better described by that of a relativistic outflow originally aspherical. In this scenario, the relativistic ejecta are detached from the freely expanding SN outflow, decelerate, spread sideways, and become spherical on a timescale of order $t_{\rm sph}\lesssim 6.2\,{\rm d}\times A^{-1}_{*}(E_{r}/10^{49}{\rm\,erg})$ \citep{Waxman2004}. The last is smaller than the timescale of the  earliest of our 2.8\,GHz observations of iPTF17cw for $\eta\gtrsim 5 $. So our discussion (which assumes a spherical blastwave) is self-consistent for $\eta \gtrsim 5$.

\subsection{Radio interpretation caveats}
Because of the relatively large distance to iPTF17cw ($d_L\approx 429$\,Mpc), our most significant radio detections of this SN are at the $\approx 5\sigma$ level. Thus, our radio measurements are affected by large errors, and so are the constraints on the spectral and temporal properties of the radio emission. Although iPTF17cw 14\,GHz counterpart has faded below detection level over the course of our follow-up, further VLA observations (in A configuration, so as to reach better sensitivity and angular resolution) are needed to confirm fading at the lower frequencies. We were not able to carry out these observations due to the fact that the VLA moved to its most compact configuration (D config) over the course of our follow-up campaign. Because our conclusion that iPTF17cw is a relativistic event relies heavily on our first measurement at 2.8\,GHz, hereafter we consider several factors that could contaminate the 2.8\,GHz flux.

In order to estimate what might be the contribution of background star formation to the measured 2.8\,GHz flux of iPTF17cw, pre-supernova photometry of the host galaxy was downloaded from the Sloan Digital Sky Survey (SDSS; \citealt{SDSS}), the Galaxy Evolution Explorer (GALEX; \citealt{GALEX}), and the Wide-field Infrared Survey Explorer (WISE; \citealt{WISE}); we also obtained images of the field from Pan-STARRS \citep{PS1} and performed aperture photometry of the host galaxy.  We then performed a simple SED fit using the \cite{BC03} stellar-population synthesis models and custom IDL software previously described in \citet{Perley+2013}.  We infer a total stellar mass of ($3.5 \pm 0.3$) $\times 10^9 M_\odot$ and a star-formation rate (averaged over the past 50 Myr) of $0.31\pm 0.05 $M$_\odot$ yr$^{-1}$, with no significant dust extinction. We note that these properties are quite similar to the LMC and to other GRB hosts at low redshift \citep[e.g.,][and references therein]{Perley2016}. The host galaxy of iPTF17cw also shows a reddish concentration at its center, which could be a bulge or weak AGN. This level of star formation implies a total 1.4\,GHz spectral luminosity of order \citep{Murphy2011}:
\begin{equation}
\left(\frac{L_{\rm 1.4\,GHz}}{\rm erg\,s^{-1}\,Hz^{-1}}\right) \approx 1.57\times10^{28}\left(\frac{\rm SFR_{\rm radio}}{M_{\odot}\,\rm yr^{-1}}\right)\approx 5\times10^{27},
\end{equation}
for the host galaxy of iPTFcw. Spectral indices of star-forming galaxies between 1.4 and 4.8 GHz are estimated to be in the range $\approx -1$ to $-0.4$ \citep{Seymour2008}. Thus, we estimate the total radio emission from the host to be $\lesssim 4\times 10^{27}$\,erg\,s$^{-1}$\,Hz$^{-1}$ at 2.8\,GHz. Our first 2.8\,GHz observation of iPTF17cw implies a radio luminosity of $\approx 10^{28}$\,erg\,$^{-1}$\,Hz$^{-1}$ \textit{concentrated in a compact region} of FWHP size $\lesssim 0.65$\arcsec, much smaller than the $R$-band size of the host galaxy (see Fig. \ref{fig:host}).  We thus consider a high level of contamination to the iPTF17cw radio counterpart from star formation in the host unlikely.  Contamination from a compact radio source, i.e. an AGN, seems unlikely considering the fact that the optical, radio, and X-ray counterparts to iPTF17cw appear to be coincident and significantly offset from the host nucleus. Confirming fading of the radio counterpart of iPTF17cw at 2.8\,GHz in future observations will strengthen  this conclusion.

We also explore possible effects related to interstellar scattering and scintillation (ISS). At the galactic coordinates of iPTF17cw ($l\approx178\,\deg$  and $b\approx42\,\deg$)  the ISS  transition  frequency  between  the strong and weak scattering regimes is $\nu_0\approx 8$\,GHz, and the Fresnel angle at the transition frequency is $\Theta_0\approx 4\,\mu$as \citep{Goodman1997,Walker2001}. This is implies that our lowest frequency observations  are  in  the  strong  scattering  regime. The modulation index in the refractive strong scattering reads: 
\begin{eqnarray}
m(\nu)\approx  \left(\frac{\Theta_0}{\Theta_s}\right)^{7/6}\left(\frac{\nu_0}{\nu}\right)^{2}~~~~{\rm for}~~\Theta_s>\Theta_r,\\
m(\nu)\approx \left(\frac{\nu}{\nu_0}\right)^{17/30}~~~~{\rm for}~~\Theta_s<\Theta_r,
\end{eqnarray}
where $\Theta_r\approx \Theta_0\left(\frac{\nu_0}{\nu}\right)^{11/5}$ and $\Theta_s$ is the source size. From the above, we see that the largest modulation index at 2.8\,GHz is $m\approx 0.55$ for a source of size $\Theta_s<\Theta_r\approx 40\,\mu$as.

To estimate whether ISS could conspire to invalidate our assumption of optically thin emission at 2.8\,GHz at 16\,d since explosion, we calculate the modulation index that would be needed to enhance potentially optically thick emission at 2.8\,GHz to a level comparable to the one observed at day 16 (so that the $2.8-5$\,GHz emission would appear optically thin, while being optically thick in reality):
\begin{equation}
m({\rm 2.8\,GHz)}\approx \frac{\sqrt{\left(f_{\rm 2.8\,GHz, 16\,d}-f_{\rm thick, 16\,d}\right)^2-(\Delta f_{\rm 2.8\,GHz, 16\,d})^2}}{f_{\rm 2.8\,GHz, 16\,d}},
\end{equation}
where $f_{\rm thick}\approx 1.5\times f_{\rm 6.2\,GHz}\times(2.8\,{\rm GHz}/6.2\,{\rm GHz})^{5/2}\approx 6\,\mu$Jy at day 16 \citep[see][for the 1.5 correction factor applied here]{Chevalier1998}; $f_{\rm 2.8\,GHz, 16\,d}$ is the measured flux at 2.8\,GHz on day 16, and $\Delta f_{\rm 2.8\,GHz, 16\,d}$ is its measurement error. We get $m({\rm 2.8\,GHz})\approx 0.85$, which is larger than the largest possible modulation index at 2.8\,GHz $m\approx 0.55$. 

We thus conclude that iPTF17cw is a plausible relativistic SN candidate. However, we stress once again that re-observing the  iPTF17cw field when the VLA will be back in its most extended configuration (March 2018), and confirming that the iPTF17cw radio counterpart has indeed faded below detection threshold at all frequencies (not only at 14\,GHz, which we have already verified), will greatly help enhance the confidence in this interpretation. 

\subsection{X-ray emission modeling}
Our Chandra detection of iPTF17cw around day 40 since GRB\,161228B implies an X-ray luminosity of $\approx (4-19)\times10^{40}$\,erg\,s$^{-1}$, which is within the range observed at a similar
epoch for the low-luminosity GRBs 980425 and 031203, most similar to the X-ray luminosity of GRB\,100316D  (Fig. \ref{fig:radio}, bottom), $\gtrsim 10\times$ larger than the X-ray luminosity of SN\,2009bb at a similar epoch \citep{Soderberg2010}, and above the X-ray luminosity upper-limit of $\approx 2.4\times10^{39}$\,erg\,s$^{-1}$ set by \textit{Chandra} observations of SN\,2012ap \citep{Margutti2014}. Because the radio luminosity of iPTF17cw is smaller than that of GRB\,980425
and GRB\,031203 (see Fig. \ref{fig:radio}), the comparable X-ray luminosity implies an flatter spectral index between the radio and X-ray bands. This is not unseen and links iPTF17cw to the low-luminosity GRBs 060218 and 100316D for which an unusually  flat radio-to-X-ray spectral index was also observed \citep[see ][for a discussion]{Margutti2013}.

From the ratio of the radio-to-X-ray (monochromatic) fluxes measured at the time of the \textit{Chandra} observation ($F_{\rm 1 keV}\approx 2\times10^{-15}$\,erg\,cm$^{-2}$\,s$^{-1}$\,keV$^{-1}$), we derive $\beta_{\rm radio-X}\sim -0.6$, which is compatible with the spectral index measured in the radio band, within the large errors (see Section \ref{radiomodel}). (We note that the extrapolation of the measured radio flux to the optical band with a similar spectral index yields an optical flux well below the one measured from iPTF17cw optical SN light curve.) The fact that the X-ray emission lies on the extrapolation from the radio band suggests a common origin of the radio and X-rays, likely  non-thermal synchrotron emission from the shocked CSM, which is usually invoked in GRB afterglows and in several radio and X-ray emitting core-collapse SNe \citep{Chevalier2006}.  We note, however, that because of the large errors that affect our radio spectral index measurement, and because of the lack of a spectral index measurement in the X-rays, we cannot exclude that, similarly to GRB\,100316D, the X-ray emission from iPTF17cw may be related to a different emission component (such as inverse Compton or prolonged central engine activity).

Within the synchrotron scenario, we can constrain the synchrotron cooling frequency to be $\nu_c(42\,{\rm d})\gtrsim 1$\,keV for iPTF17cw, due to the apparent lack of a spectral steepening between the radio and X-rays. Using Equation (11) of \citet{Chevalier2006} (where we correct the normalization constant, erroneously reported there as $10^{10}$\,GHz to $10^{10}$\,Hz):
\begin{equation}
\nu_{c}  \approx8\times10^{12}\left(\frac{\epsilon_B}{0.1}\right)\alpha^{-3/2}A_*^{-3/2}\left(\frac{t}{\rm 42\,d}\right)~{\rm Hz}.
\end{equation}

From the above it is evident that $\nu_c\gtrsim 1$\,keV would require an unusually low value of $A_*$. The lack of a spectral steepening at high frequencies is common to several other SN including the radio-loud BL-Ic PTF11qcj \citep{Corsi2014}, SN\,2009bb, and SN\,1998bw, and may be explained by invoking a cosmic-ray dominated SN shock \citep{Ellison2000,Chevalier2006}. 

We consider in what follows other possible sources of X-rays generally invoked in the study of stripped-envelope core-collapse SNe \citep{Chevalier2006}, such as thermal emission associated with the shock-heated ejecta, and  inverse Compton (IC) up-scattering of optical photons by radio-emitting electrons. 

As discussed in \citet{Corsi2014}, within the thermal emission hypothesis, X-rays are produced while  the  forward  shock  plows  into  the  CSM  and /or  by  the reverse  shock  heating  the  ejecta.  Thus,  the  X-ray  luminosity depends  on  the  density  of  the  emitting  material,  which  cools by free-free emission. Because the X-ray luminosity is directly proportional to the square of the mass-loss rate \citep[see e.g. Eq. (16) in][]{Corsi2014}, for an event like iPTF17cw ($\approx 10^{-6}\,M_{\odot}\,{\rm yr}^{-1}$) we don't expect free-free emission to be the dominant contributor to the iPTF17cw X-ray counterpart detected by \textit{Chandra}.

The IC mechanism is expected to be most important on timescales comparable to the optical SN maximum, since the ratio of flux measured in the radio and X-ray bands is given by:
\begin{equation}
\frac{F_X}{\nu_{\rm radio} F_{\nu, \rm radio}}\sim \frac{U_{ph}}{U_B}
\end{equation}
where $U_{\rm ph}$ is the energy density in up-scattered photons, which can be estimated as \citep{Bjornsson2004,Soderberg2006}:
\begin{equation}
U_{\rm ph}\approx 0.4 \times \left(\frac{L_{\rm bol}}{10^{42}\,{\rm erg\,s}^{-1}}\right)\left(\frac{t}{1\,\rm d}\right)^{-2} \left(\frac{{\rm v}_s}{c}\right)^{-2}~{\rm erg\,cm^{-3}}.
\end{equation}
For iPTF17cw, even setting $L_{\rm bol}$ to its peak value of $L_{\rm bol,peak}\approx 10^{43}$\,erg\,s$^{-1}$ (at $t\approx 18$\,d since GRB\,161228B), and using v$_{s}\approx 0.85\,c$ (see Section \ref{radiomodel}), we estimate $U_{\rm ph}\lesssim 2\times 10^{-2}$\,erg\,cm$^{-3}$. From our radio observations, we also derive $\nu_{\rm radio} F_{\nu, \rm radio} \lesssim 2\times 10^{-18}$\,erg\,cm$^{-2}$\,s$^{-1}$. Setting $B\sim 0.3 (t/16\,{\rm d})^{-1}$\,G based on what we derived in Section \ref{radiomodel} and assuming a typical $B\propto t^{-1}$ evolution of the magnetic field \citep{Chevalier1998}, we derive the conservative upper-limit:
\begin{equation}
F_X (42\,{\rm d})\lesssim  10^{-16}~{\rm erg\,cm^{-2}\,s^{-1}},~~
\end{equation}
which is a factor of $\approx 10$ below the X-ray flux measured by \textit{Chandra} at $t\approx 42$\,d since GRB\,161228B.

We thus conclude that a synchrotron origin from a cosmic-ray dominated shock is a more likely scenario for the radio and X-ray emission of iPTF17cw. However (as discussed above) we cannot exclude other mechanisms for the origin of the X-ray emission, such as long-lasting
activity of the explosion central engine \citep[e.g., a black hole plus torus system, or a magnetar progenitor; ][]{Margutti2013}. Modeling of iPTF17cw emission within these more complex scenarios is (given our limited dataset) beyond the scope of this paper.
\section{Summary and conclusion}
\label{conclusion}
We have presented the discovery, classification, and radio-to-X-ray follow-up observations
of iPTF17cw, a BL-Ic SN discovered by the iPTF while carrying out an extensive follow-up campaign of the (unrelated) Advanced LIGO event  GW170104. iPTF17cw spectroscopic and photometric properties show similarities with the engine-driven SN\,1998bw and the relativistic SN\,2009bb. We estimate a Nickel mass similar to that of SN\,1998bw, while the total energy of the ejecta and ejecta mass are more similar to that of SN\,2009bb.

Our radio data suggest that iPTF17cw is a new member of the rare sample of relativistic SNe discovered independently of a $\gamma$-ray trigger. Given the temporal and spatial compatibility of iPTF17cw with GRB\,161228B, we also consider iPTF17cw a potential new member of the limited sample of GRB-associated (engine-driven) BL-Ic SNe. We caution, however, that our radio observations are affected by large errors (given the large distance to iPTF17cw). Thus, further late-time VLA follow-up observations with high resolution are warranted to confirm the relativistic nature of iPTF17cw.

Radio-to-X-ray follow-up observations of iPTF17cw enabled us to estimate the properties of the fastest moving ejecta and of the CSM. The radio luminosity of iPTF17cw is $\approx 10$ times smaller that SN\,1998bw, but comparable to that of GRB\,100316D. The fraction of energy coupled to the fastest emitting material is $\gtrsim 10$ times smaller than in SN\,1998bw, while the CSM density around iPTF17cw is similar to that found around other relativistic BL-Ic SNe such as SN\,2009bb, and consistent with values inferred for long GRBs. The relatively bright X-ray emission is compatible with the extrapolation from the radio band within a synchrotron emission scenario, and the implied X-ray luminosity is most similar to that of GRB\,100316D. 

Thanks to the much increased rate of iPTF discoveries of BL-Ic SNe, we have just now begun to observationally constrain the fraction $f$ of BL-Ic SNe that may be powered by central engines. Considering the size of the sample of BL-Ic SNe with radio observations we have collected so far, the discovery of iPTF17cw independently of a $\gamma$-ray trigger fits our expectations for $30\%\lesssim f\lesssim 40\%$ \citep{Corsi2016,Podsiadlowski2004}. This result also demonstrates that our discovery rate of engine-driven BL-Ic is likely to be substantially boosted by the advent of the Zwicky Transient Facility \citep[ZTF;][]{Smith2014,Bellm2016}. In the ZTF era, we expect to collect in one year a sample of BL-Ic SNe with radio observations as large as the one we collected with the iPTF in five years, and the discovery of events like iPTF17cw will be greatly facilitated.\\
\newline

\acknowledgments \small
A.C. and N.T.P. acknowledge support from the National Science Foundation under CAREER Grant No. 1455090. A.C. and N.T.P. also acknowledge partial support from the 
\textit{Swift} Cycle 12 GI program (Grant No. NNX17AF93G). Support for this work was in part provided by 
the National Aeronautics and Space Administration through Chandra Award No. 19500451 issued 
by the Chandra X-ray Observatory Center, which is operated by the Smithsonian Astrophysical Observatory 
for and on behalf of the National Aeronautics Space Administration under contract NAS8-03060. 
This work was also supported by the GROWTH project funded by the National Science Foundation under Grant No. 1545949.
The National Radio Astronomy Observatory is a facility of the National Science Foundation 
operated under cooperative agreement by Associated Universities, Inc.
These results made use of the Discovery Channel Telescope 
at Lowell Observatory. Lowell is a private, non-profit institution 
dedicated to astrophysical research and public appreciation of astronomy 
and operates the DCT in partnership with Boston University, the University 
of Maryland, the University of Toledo, Northern Arizona University and 
Yale University.  The upgrade of the DeVeny optical spectrograph has been 
funded by a generous grant from John and Ginger Giovale.
This work made use of data supplied by the UK Swift Science Data Centre 
at the University of Leicester. This work is partially based on observations collected with the Liverpool Telescope, which is operated on the island of La Palma by Liverpool John Moores University in the Spanish Observatorio del Roque de los Muchachos of the Instituto de Astrofisica de Canarias with financial support from the UK Science and Technology Facilities Council.
The Pan-STARRS1 Surveys (PS1) have been made possible through contributions of the Institute for Astronomy, the University of Hawaii, the Pan-STARRS Project Office, the Max-Planck Society and its participating institutes, the Max Planck Institute for Astronomy, Heidelberg and the Max Planck Institute for Extraterrestrial Physics, Garching, The Johns Hopkins University, Durham University, the University of Edinburgh, Queen's University Belfast, the Harvard-Smithsonian Center for Astrophysics, the Las Cumbres Observatory Global Telescope Network Incorporated, the National Central University of Taiwan, the Space Telescope Science Institute, the National Aeronautics and Space Administration under Grant No. NNX08AR22G issued through the Planetary Science Division of the NASA Science Mission Directorate, the National Science Foundation under Grant No. AST-1238877, the University of Maryland, and Eotvos Lorand University (ELTE).

\bibliographystyle{aasjournal}
\bibliography{Cors515}
\newpage

\begin{center}
\begin{longtable}{lcccc}
\caption{Optical photometry of iPTF17cw. \label{phot_tab}}
\\
\hline
\hline
Date & Telescope & Band & Magnitude \\
(MJD) & &  & [AB]\\ 
\hline
\endhead
\hline
57760.291& P48 & $R$ & $19.46\pm0.10$\\
57760.319 & P48 & $R$ &  $19.557\pm0.072$\\
\hline
57768.305 & P60 & $g$ & $19.041\pm 0.077$\\
57770.194 & P60 & $g$ & $19.235\pm0.034$\\
57771.149 & P60 & $g$ & $19.294\pm0.054$\\
57780.219 & P60 & $g$ & $20.54\pm0.17$\\
57783.277 & P60 & $g$ &  $>20.6$\\
57787.103 & P60 & $g$ &   $>17.6$\\
57789.186 & P60 & $g$ & $20.93\pm0.16$\\
57793.187 & P60 & $g$ &   $>19.8$\\
\hline
57768.301 & P60 & $r$ & $18.703 \pm 0.038$ \\
57770.191 & P60 & $r$ & $18.721 \pm 0.022$ \\
57771.146 & P60 & $r$ & $18.793\pm  0.032$\\
57783.273 & P60 & $r$ & $19.446\pm 0.045$\\
57784.124 & P60 & $r$ & $>18.4$ \\
57789.183 & P60 & $r$ & $19.998 \pm 0.084$\\
57793.184 & P60 & $r$ & $>18.3$\\
57794.123 & P60 & $r$ &$>19.4$\\
57797.444 & P60 & $r$ & $20.14 \pm 0.19$\\
57799.108 & P60 & $r$ & $20.63\pm 0.17$\\
\hline
57768.303 & P60 & $i$ & $18.901\pm 0.066$\\
57770.193 & P60 & $i$ & $18.872\pm 0.040$\\
57771.147 & P60 & $i$ & $18.902\pm 0.053$\\
57780.217 & P60 & $i$ & $19.142\pm 0.060$\\
57783.275 & P60 & $i$ & $19.223\pm 0.067$\\
57789.185 & P60 & $i$ & $19.692\pm 0.062$\\
57793.185 & P60 & $i$ & $>19.0$\\
57794.125 & P60 & $i$ & $>18.1$\\
57797.446 & P60 & $i$ & $20.17 \pm 0.17$\\
57799.110& P60 & $i$ &  $>20.3$\\
\hline
\end{longtable}
\end{center}

\begin{center}
\begin{longtable}{ccccc}
\caption{Radio observations of iPTF17cw. Epochs are calculated since the \textit{Fermi}/GBM trigger time of GRB\,161228B (MJD 57750.552). \label{radioTab}}
\\
\hline
\hline
MJD & Epoch & Instr.:config & Freq. & Flux Density \\
          &   (d)&    & (GHz)    & ($\mu$Jy/beam)  \\
\hline
\endhead
57763.140 &12.6 &  VLA:A & 6.2 & $38.1\pm7.3$\\
\hline
57766.260 & 15.7 & VLA:A & 6.3 & $30.4\pm5.9$\\
\hline
57766.464 & 15.9 & VLA:A &  2.8  & $50\pm10$  \\
 ''& '' &  " &14   & $25.4\pm6.2$ \\
\hline
 57772.115 & 21.6 & VLA:A & 6.2 & $19.9\pm6.2$\\
 \hline
 57775.451 & 24.9 & VLA:A & 2.8 & $41.4\pm8.7$\\
 "                 &   "     &  "        & 14  & $21.1\pm6.3$\\
 \hline
 57781.217 & 30.7 & VLA:AnD & 6.2 & $22.4\pm5.5$\\
 \hline
57792.182 & 41.6 & VLA:AnD & 2.8 & $44\pm14$\\
" & " & " & 6.2 & $19\pm20$ \\
\hline
 57816.103& 65.6 & VLA:D & 14 & $20.2\pm6.8$\\
\hline
57855.470& 105 & VLA:D & 14 & $10.7\pm5.2$\\
\hline
\end{longtable}
\end{center}
\begin{center}
\begin{longtable}{lcccccc}
\caption{X-ray observations of iPTF17cw. Epochs are calculated since the \textit{Fermi}/GBM trigger time of GRB\,161228B (MJD 57750.552).\label{X}}
\\
\hline
\hline
Date & Epoch & Instrument & Band & Exp. Time & Count Rate & Flux (unabs)\\
(MJD) & (d) & (keV) & (ks) & (s$^{-1}$) & (erg\,cm$^{-2}$\,s$^{-1}$)\\ 
\hline
\endhead
57766.757 & 16.2 &\textit{Swift}-XRT &  0.3-10 & 1.4 & $< 8.1 \times 10^{-3}$ & $< 3.4 \times 10^{-13}$\\
57771.524 & 21.0&\textit{Swift}-XRT & 0.3-10 & 7.5 & $< 2.0 \times 10^{-3}$ & $< 8.5 \times 10^{-14}$\\
57792.390 & 41.8&\textit{Chandra}-ACIS & 0.5-7  & 9.86  & $\left(5.0_{-2.9}^{+4.8}\right) \times10^{-4}$ & $\left(5.0_{-2.9}^{+4.7}\right) \times 10^{-15}$\\
\hline
\end{longtable}
\end{center}

\end{document}